\documentclass[%
 aps,
 pre,%
 amsmath,amssymb,amsfonts,
 reprint,%
]{revtex4-1}
\usepackage{graphics,graphicx}
\usepackage{cleveref}
\usepackage{color}
\usepackage{tikz}
\usetikzlibrary{decorations.pathreplacing}
\usepackage{array}
\usepackage{droidsans}
\usepackage{charter}
\usepackage[T1]{fontenc}

\usepackage{setspace}
\usepackage[compact]{titlesec}
\usepackage{subcaption}
\usepackage{amsmath}
\usepackage{amsmath,bm}
\usepackage{graphicx}
\usepackage{adjustbox}
\usepackage{float}
\usepackage{morefloats}
\usepackage{cleveref}
\usepackage{color}
\usepackage{soul}

\definecolor{cream}{RGB}{222,217,201}
\definecolor{forestgreen}{rgb}{0.13, 0.55, 0.13}
\definecolor{brown}{rgb}{0.24, 0.17, 0.12}
\definecolor{magenta}{rgb}{1.0, 0.0, 1.0}
\usepackage{morefloats}

\newcommand{\mi}{\boldsymbol{-} \mathrel{\mkern -16mu} \boldsymbol{-}}

\def\.{\cdot}
\def\l{{\lambda}}

\def\be{{\begin{equation}}}
\def\ee{{\end{equation}}}

\def\l[{{[\![}}
\def\r]{{]\!]}}

\definecolor{forestgreen}{rgb}{0.13, 0.55, 0.13}
\definecolor{brown}{rgb}{0.24, 0.17, 0.12}

\begin{document}


\title[]{Large deformation electrohydrodynamics of a Skalak elastic capsule in AC electric field}
\author{Sudip Das, Rochish M Thaokar}
 \email{rochish@che.iitb.ac.in}
\affiliation{Department of Chemical Engineering,
  Indian Institute of Technology Bombay, Mumbai 400 076, India}
\date{\today}

\begin{abstract}

The axisymmetric electrohydrodynamic deformation of an elastic capsule with capacitive membrane obeying Skalak law under uniform AC electric field is investigated using analytical and boundary integral theory. The low capillary number (the ratio of destabilizing shear or electric force to the stabilizing elastic force) regime shows time-averaged prolate and oblate spheroid deformations, and the time-periodic prolate-sphere, oblate-sphere breathing modes are commensurate with the time averaged-deformation. A novel prolate-oblate breathing mode is observed due to an interplay of finite membrane charging time and the field reversal of the AC field. The study, when extended to high capillary number, shows new breathing modes of cylinder-prolate, cylinder-oblate, and biconcave-prolate deformation. These are the results of highly compressive normal Maxwell stress at the poles and are aided by weak compressive equatorial stress, characteristic of a capacitive membrane. The findings of this work should form the basis for the understanding of more complex biological cells and synthetic capsules for industrial applications.     

\end{abstract}
\maketitle

\section{Introduction}
Elastic capsules and vesicles consist of a thin membrane separating the inner and outer fluids and have been commonly used as bio-mimetic systems to understand the mechanics of biological cells~\cite{helfrich73,helfrich73a}, which are much more complex. Elastic capsules form good models for cells such as red blood cells. Moreover, they are used in technological applications such as microreactors~\citep{yang06} and diagnostics applications~\citep{yang96,xiong13}. Therefore, exploring the mechanical and rheological behavior of an elastic capsule is an important topic of research~\cite{brondsted87,lensen08,tanner11}.

Vesicles consist of a lipid bilayer membrane which resists deformation by bending stress~\cite{helfrich89,stephen86}. The lipid bilayers are characterized by negligible shear elasticity but are area conserving~\cite{helfrich74,seifert97}, which leads to the generation of a non-uniform but isotropic tension in the membrane corresponding to the imposed stresses. The membrane in elastic capsules, on the other hand, is made up of a crosslinked network of proteins (as in red blood cells) or other surface active molecules.  Elastic capsules have a finite shear elasticity and they resist deformation by in-plane shear and dilatational stresses, these are related to the in-plane shear and dilatational strains~\cite{barthesbiesel02}. The elasticity (shear and dilatation) can generate non-uniform, anisotropic tensions in the membrane. There are other differences between a capsule and a vesicle, for example, while an area preserving spherical vesicle cannot deform, a spherical elastic capsule with a compressible membrane can admit deformations. 

The electrohydrodynamics of vesicles have been fairly well investigated in the last few decades in the context of cell electroporation~\cite{crowley73,tieleman04}, electrofusion as well as cell dielectrophoresis~\cite{neumann89,thomas93,ronald96} and electrodeformation~\cite{veerapaneni16,ebrahim15,ebrahim15a,mcconnell13,mcconnell15sm,mcconnell15}. The electric field models too, for a vesicle, have evolved over the years. While Helfrich modeled the electrostatics of a vesicle subjected to an electric field by assuming a dielectric drop suspended in a conducting fluid~\citep{helfrich74,helfrich91}, there have been attempts to model it as a leaky dielectric drop in a leaky dielectric fluid~\cite{helfrich88}. Recently a more realistic capacitor model for electrostatics~\cite{katherine99,vlahovska12,vlahovska13,ebrahim15a,veerapaneni16} have been introduced.
 The capacitor model for electrostatics assumes the vesicle to be made up of an infinitesimally thin dielectric layer characterized by a membrane capacitance and conductance enclosing a leaky dielectric fluid and embedded in another leaky dielectric fluid. Several timescales can then be identified apart from the hydrodynamic timescale. These associated timescales are the charge relaxation timescale of the outer fluid, the time required for the two fluids to behave like leaky dielectrics, namely the Maxwell-Wagner relaxation time and the membrane charging time~\cite{mcconnell13}.

AC fields are typically preferred in operations such as electro-deformation and dielectrophoresis involving vesicles and cells. An advantage of AC fields, apart from preventing electrolysis of the solution, is the control of trans-membrane potential in such cells and vesicles~\citep{grosse92}, thereby preventing electroporation~\cite{marszalek90} and enabling systematic investigations. Experiments and analytical theories on electrohydrodynamics of vesicles in the low electric field limit show that the variation of deformation with frequency sensitively depends upon the conductivity ratio of the internal and external fluid media~\cite{vlahovska09,peterlin10}. This dependence is due to an interplay of the different electric timescales and the applied frequency. For more conducting inner medium, prolate spheroids and spherical steady shapes are observed, while in the case of more conducting outer medium, spherical, oblate and prolate spheroidal shapes are observed. At high electric fields, experimental investigations~\cite{karin06,dimova07} showed squaring of a vesicle in a DC field which was reaffirmed and explained by a few 2-D \cite{mcconnell15,mcconnell13} and 3-D~\cite{veerapaneni16,ebrahim15a,ebrahim15} computational analyses. Further studies showed that unlike DC fields, a vesicle subjected to high AC electric field showed deformation modes such as breathing and oscillating~\cite{mcconnell15sm}. 

Theoretical analyses of capsule~\citep{barthesbiesel80,barthesbiesel81,BARTHESBIESEL81a,jong00,kessler09,rt16,seifert11,vlahovska11a} in the limit of small deformation, indicate a linear relationship between the extent of deformation and the capillary number. In this limit, other aspects of capsule deformation such as orientation, tank-treading motion, and rheology of dilute suspension have also been reported. Several computational analyses of a capsule in external fluid flow in the large deformation limit present the dependence of the deformation on the different membrane constitutive laws~\citep{barthesbiesel91,lac04} and their break up (of a capsule with strain softening membrane) beyond a critical capillary number~\citep{barthesbiesel91,pozrikidis95,chang93,barthesbiesel00,loubens15} 

Relatively few studies have been reported on the deformation of elastic capsules under DC~\citep{jong00} and AC~\citep{rt16} electric fields. These studies were restricted to small deformation limit. A strong influence of the applied electric field on the deformation of the capsule was reported in the experimental, theoretical and computational analysis of a conducting capsule enclosing a conducting fluid and suspended in a dielectric fluid medium~\citep{karyappa14}. Moreover, this analysis suggests the usability of the electrohydrodynamics to determine the linear and nonlinear elastic properties of a capsule. A comprehensive analysis of the deformation of an elastic capsule with a dielectric membrane in DC electric field is reported in our earlier work~\citep{sudip17} which shows several complex modes of deformation, such as biconcave and hexagonal shapes. These new modes of deformation are not observed in the deformation of vesicles in DC electric field~\cite{veerapaneni16,ebrahim15a,ebrahim15,mcconnell15sm}.

In this analysis, we investigate the deformation of an elastic capsule in strong AC electric field, wherein an interaction between the electric stress and restoring elastic forces could lead to results different than those obtained for vesicles. Also, the phase diagrams are constructed for deformation of a spherical elastic capsule in low (using both the boundary integral calculation and analytical theory) and high (using boundary integral method) capillary numbers in AC electric field and attempts are made to understand the underlying physics. The capacitor model is used for modeling the electrostatics~\citep{mcconnell15sm,mcconnell15,mcconnell13,Hu16,veerapaneni16,ebrahim15,ebrahim15a} while the capsule is modeled as a strain hardening Skalak membrane~\citep{skalak73}. The deformation of an object in AC field has time-averaged and time-periodic parts, and their dependence on frequency could be significantly different. The physics of both these kinds of deformation are analyzed in this work. 

\section{Problem description}
\begin{figure}
\centering
    \includegraphics[width=0.4\textwidth]{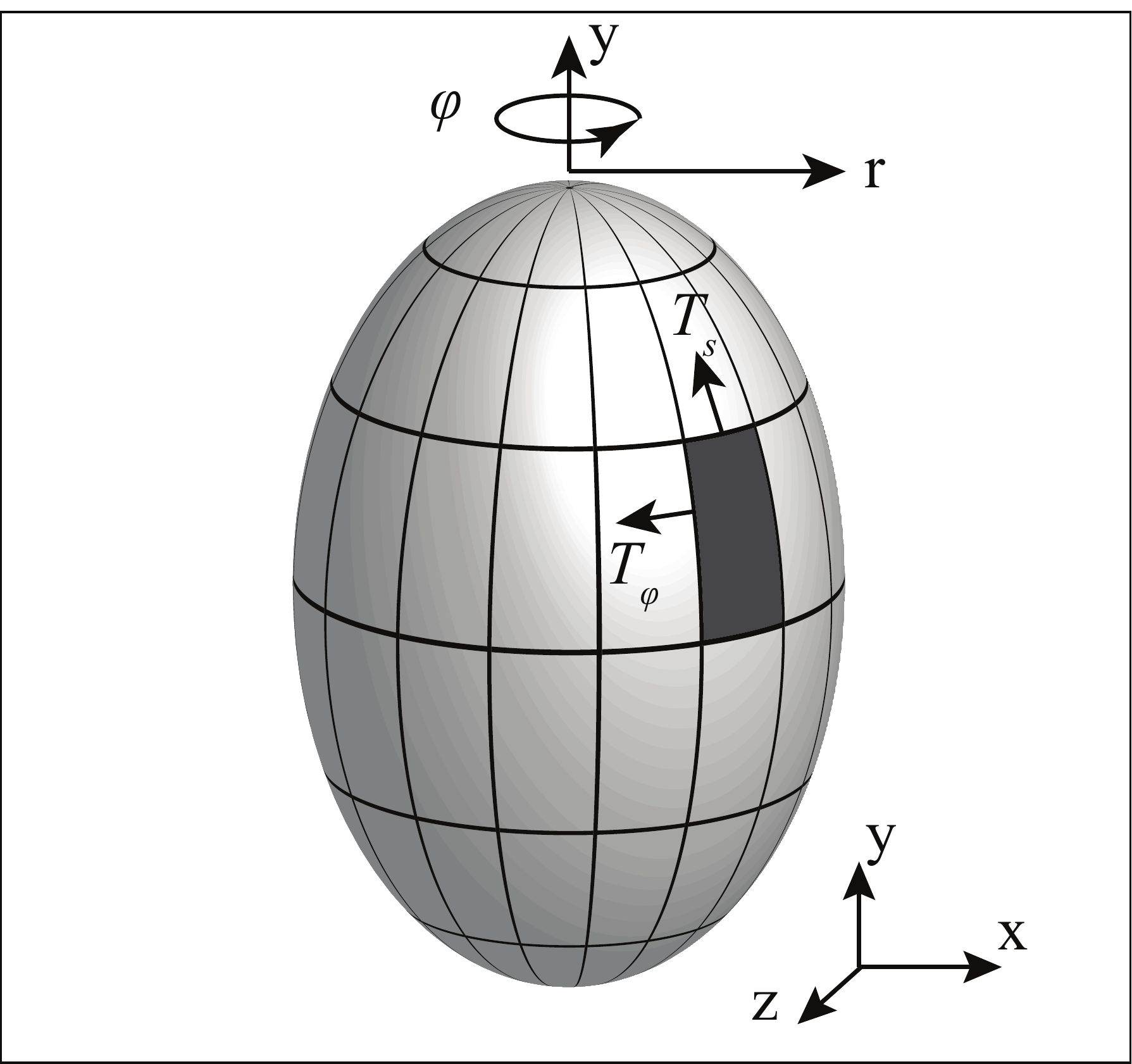}   
\caption{Schematic representation of an axisymmetrically deformed elastic capsule. The electric field is applied parallel to the $y$-axis. $T_s$ and $T_\phi$ are meridional and azimuthal membrane tensions, respectively. }
\label{fig:schematic}
\end{figure}

A spherical capsule is placed in an applied uniform AC electric field in which the direction of the applied field is parallel to the axis of symmetry, the $y$-axis (shown in~\cref{fig:schematic}). The AC electric field is defined by $\tilde{{\bf E}}^{\infty}_{AC}=\tilde{E_0}\cos(\tilde{\omega} \tilde{t}) {\bf e}_y$, where $\tilde{E_0}$, and $\tilde{\omega}$ are the intensity and frequency of the applied electric field, respectively and $\tilde{t}$ is the time. 

Both the inner and outer fluid media are Newtonian with the outer fluid viscosity $\mu$ and inner fluid viscosity $\lambda\mu$, where $\lambda$ is the viscosity ratio. Conductivity and dielectric constant of the fluid media are $\sigma_{i,e}$ and $\epsilon_{i,e}$, respectively. In this work, subscripts $i$ and $e$ stand for inner and outer fluid media, respectively. Electrical properties used in the computations are described as ratios with respect to the outer fluid media: $\sigma_r=\sigma_i/\sigma_e$ and $\epsilon_r=\epsilon_i/\epsilon_e$. The membrane is considered to be purely elastic with elastic modulus $E_s$ and it provides resistance to bending deformation proportional to the bending rigidity $\kappa_b$. The membrane electrical properties are capacitance $C_m$ and conductance $G_m$.

The charge relaxation time of the external fluid, ${\epsilon_e\epsilon_0}/{\sigma_e}$, where $\epsilon_0$ is the permittivity of the free space, is considered as the scaling parameter for time and the velocity of the fluid is scaled by $E_s/\mu_e$. Frequency of the applied electric field is scaled by ${\sigma_e}/{\epsilon_e\epsilon_0}$, the inverse of the parameter used for scaling of time. The intensity of the electric field (${\bf \tilde E}$), potential ($\tilde \phi$), and stress are scaled by  $E_0$, $E_0 a$, and $E_s/a$, respectively, where $a$ is the radius of the capsule. All the variables with and without ($\ \tilde{}\ $) represent dimensional and scaled quantities, respectively. In this analysis, the magnitude of the destabilizing force due to the electric field is expressed relative to the stabilizing elastic force by the nondimensional capillary number,  $Ca=a \epsilon_e\epsilon_0E_0^2/E_s$.

For the simplicity of the calculations, a very thin (assumed to be two dimensional) membrane is considered to be impermeable to both the fluids, thereby keeping the volume of the capsule constant. The effect of the bending rigidity is considered to be important as it prevents the formation of sharp corners as well as helps in computation by averting numerical instabilities~\cite{dodson09}.

The resultant of the hydrodynamic force and non-hydrodynamic forces due to electric Maxwell stress, elastic force, and force because of resistance to bending cause a change in the shape of a capsule. Stokes equations are considered as the model equations to study the capsule hydrodynamics~\citep{rallison78,poz92}, electric stresses are calculated by solving Laplace's equation along with normal electric current continuity~\cite{mcconnell15sm,mcconnell15,mcconnell13,ebrahim15,ebrahim15a}. Membrane elastic forces are determined from the well known Skalak law for the strain hardening membrane~\cite{skalak73}. Bending force is a function of up to fourth order derivative of the position of the interface, and it is calculated using the spectral method~\citep{kwak01}. 

\section{Problem formulation}
The deformation of a capsule in an AC field is studied analytically using the small deformation theory as well as numerically, using axisymmetric boundary integral method to understand large deformations. 

\subsection{Boundary integral formulation}
The elastic force, electric force, the force due to bending and the  hydrodynamic forces are discussed in detail in the following sections. 
\subsubsection{Membrane elasticity}
Nonlinear response of a $2D$ membrane to deformation is modeled by different membrane constitutive laws, namely Generalized Hooke's law, Skalak law, Mooney-Rivlin law, Evan-Skalak law etc., and can be broadly categorized into strain hardening membrane models and strain softening membrane models. In this analysis, the deformation of a capsule is investigated using the Skalak model which exhibits a strain hardening behavior.  According to the Skalak law~\cite{skalak73}, the tension in the principal direction $i$ (the other principal direction being $j$), can be obtained, in terms of principal extension ratios, $\lambda_{i,j}$, as
\begin{equation}\label{eq:skalaknd}
\begin{aligned}
{T}_{i}^{SK} &=G^{SK} \frac{1}{\lambda_i\lambda_j}\left[\lambda_i^2(\lambda_i^2-1)+C(\lambda_i\lambda_j)^2\{(\lambda_i\lambda_j)^2-1\}\right],
\end{aligned}
\end{equation}
where $C$ is the area dilatation parameter for Skalak membrane, larger the value of $C$, stiffer is the membrane. In this analysis, $C=1$ is considered. The first term, on the right-hand side of~\cref{eq:skalaknd},  is corresponding to the shear deformation with the shear modulus $G^{SK}=E_s(1+C)/2(1+2C)$ and the second term accounts for the area dilatation with the dilatation modulus $CG^{SK}$. Membrane tension in the principal direction $j$ can be obtained by interchanging indices. In our analysis, principal directions $i$ and $j$ correspond to the meridional and azimuthal directions, respectively and the corresponding tensions (force/length) are considered to be $T_s$ and $T_\phi$. These tensions contribute to the normal and tangential elastic tractions (force/area) as
\begin{align}
 f_n^{el} &= -(K_sT_s+K_{\phi}T_\phi)\\
 f_t^{el} &= \frac{d T_s}{ds}+\frac{1}{r}\frac{dr}{ds}(T_s-T_\phi),
 \end{align}
such that overall elastic traction acting at the interface is ${\bf f}^{el}=f_n^{el}{\bf n}+f_t^{el}{\bf t}$, where ${\bf n}$ and ${\bf t}$ are the outward normal and clockwise tangent vectors, respectively. The arc length in the meridional plane, $s$, is measured from the north pole to the south. The principal curvatures are defined as
\begin{equation}\label{curve}
 K_s=\left|\frac{d{\bf t}}{ds}\right| \quad  K_\phi=\frac{n_r}{r},
\end{equation}
and the components of the unit tangent and the unit normal in $r$ and $y$ directions are calculated as
\begin{equation}\label{normtan}
t_r=-n_y=\frac{dr}{ds} \quad t_y=n_r=\frac{dy}{ds}.
\end{equation}

 \subsubsection{Bending resistance}
Although the bending resistance of a capsule with a very thin elastic membrane  is small compared to the elastic stresses, it can contribute significantly, especially in regions of high curvature and is also known to  prevent numerical elastic instabilities~\cite{dodson09} (arising from the large compressive stresses). 

The dimensionless traction due to bending is given by
\begin{equation}
 {\bf f}^b =\hat \kappa_b \left[2{ \Delta_s} H+4 H(H^2-K_G)\right] {\bf n},
\end{equation}
where, $H$ and $K_G$ are mean curvature and Gaussian curvature, respectively,
\begin{equation}
 H=\frac{1}{2}(k_s+k_\phi) \text{\ and,\ } K_G=k_sk_\phi.
\end{equation}
Nondimensional bending rigidity, $\hat \kappa_b=\kappa_b/a^2E_s$, is the bending rigidity modulus relative to the elastic modulus. Laplace Beltrami of the mean curvature in cylindrical coordinate system is defined as 
\begin{equation}
 \Delta_sH=\nabla_s\cdot(\nabla_sH)=\frac{1}{r\mid{\bf X}_s\mid}\frac{\partial}{\partial s}\left(\frac{r}{\mid{\bf X}_s\mid}\frac{\partial H}{\partial s}\right),
\end{equation}
where $\mid{\bf X}_s\mid=\sqrt{\left(\frac{\partial r}{\partial s}\right)^2+\left(\frac{\partial y}{\partial s}\right)^2}$~\cite{hu14}. The spectral method is used to calculate higher order derivatives for the calculation of Laplace Beltrami of mean curvature~\cite{trefethen94}.  

\subsubsection{Electrostatics}
In an externally applied electric field $E_0\cos(\omega t)$, in the absence of any interacting particle, the potential is given by $\phi^{\infty}=-y\cos\omega t$. The internal and the external potential of the capsule satisfy Laplace equation, $\nabla^2\phi_{i,e}=0$, where $\phi_i$ and $\phi_e$ are internal and external potentials. Using Green's theorem, the solution of the Laplace equation can be obtained in integral forms as~\cite{mcconnell15sm,mcconnell15,mcconnell13} 
\begin{widetext}
\begin{align}
 \frac{1}{2}\phi_i({\bf x}_0) &= \int_s \left[G^E({\bf x},{\bf x}_0){\bf \nabla}\phi_i({\bf x})\cdot {\bf n}({\bf x})-\phi_i({\bf x}){\bf n}({\bf x})\cdot {\bf \nabla}G^E({\bf x},{\bf x}_0)\right] dS({\bf x}),\label{eq:pot1}\\ 
 \frac{1}{2}\phi_e({\bf x}_0) &= \phi^\infty ({\bf x}_0)-\int_s \left[G^E({\bf x},{\bf x}_0){\bf \nabla}\phi_e({\bf x})\cdot {\bf n}({\bf x})-\phi_e({\bf x}){\bf n}({\bf x})\cdot {\bf \nabla}G^E({\bf x},{\bf x}_0)\right] dS({\bf x})\label{eq:pot2},
\end{align}
 \end{widetext}
where $G^E({\bf x}_0,{\bf x})=\frac{1}{4\pi |\hat{{\bf x}}|}$ is the Green's function for the Laplace's equation in three-dimensional free space. Here, $dS({\bf x})$ is the differential surface area,  ${\bf x}_0$ and ${\bf x}$ are the source point (singular point) and the load point, respectively and $\hat{{\bf x}}={\bf x}-{\bf x}_0$.  
Transmembrane potential is defined as the discontinuity of potential across the interface, given by
\begin{equation}\label{eq:phim}
\phi_m=\phi_i-\phi_e. 
\end{equation}
Electrical current continuity across the interface is 
\begin{equation}\label{eq:currentcont}
 \sigma_r E_{n,i}+\epsilon_r\frac{dE_{n,i}}{dt}=E_{n,e}+\frac{dE_{n,e}}{dt}=\hat C_m \frac{d \phi_m}{dt}+ \hat G_m \phi_m
\end{equation}
where $\hat C_m={a {C_m}}/{\epsilon_e\epsilon_0}$ and $ \hat G_m={a {G}_m}/{\sigma_e}$ are the dimensionless capacitance and conductance of the membrane of the capsule, respectively. Solving \cref{eq:pot1,eq:pot2} with the definition of transmembrane potential (\cref{eq:phim}) and current continuity (\cref{eq:currentcont}), the transmembrane potential and the internal and external normal and tangential electric fields (defined as $-\partial\phi/\partial s$) can be calculated. 

The traction associated with the Maxwell electric stress acting at the interface is ${\tilde {\bf f}^E}=  (\tilde {\bf T}^E_e-\tilde {\bf T}^E_i)\cdot{\bf n}=\tilde\tau^{E}_n{\bf n}+\tilde\tau^{E}_t{\bf t}$, where the stress tensor $\tilde{\bf T}^E=\epsilon \epsilon_0 \left(\tilde {\bf E}\tilde{\bf  E}-\frac{1}{2} \tilde E^2{\bf I} \right)$, ${\bf I}$ is the identity tensor. The components of the electric stress in dimensionless form are obtained as
 \begin{align}
   \tau_n^E &= \frac{1}{2}\left[(E_{n,e}^2-E_{t,e}^2)-\epsilon_r(E_{n,i}^2-E_{t,i}^2)\right],\\
 \tau_t^E &= E_{n,e}E_{t,e}-\epsilon_rE_{n,i}E_{t,i},
 \end{align}
and the components of the electric traction can be obtained as 
\begin{equation}
 f_y^E=\tau_t^Et_y+\tau_n^En_y \quad \text{and}\quad f_r^E=\tau_t^Et_r+\tau_n^En_r.
\end{equation}

\subsubsection{Hydrodynamics}
The typical size of capsules used in applications is $\sim$\,$100\mu m$ indicating that hydrodynamics can be described as a low Reynolds number phenomena. Therefore, considering Stokes equations to describe the hydrodynamics, Green's theorem can be used to obtain a boundary integral equation given by 
\begin{widetext}
\begin{equation}\label{eq:veleqn}
{\bf u}({\bf x}_0) =-\frac{1}{1+\lambda}\frac{1}{4\pi}\int_s\Delta{{\bf f}}({\bf x})\cdot{\bf G}({\bf x},{\bf x}_0)dS({\bf x})+\frac{1}{4\pi}\frac{1-\lambda}{1+\lambda}\int_s {\bf u}({\bf x})\cdot{\bf Q}({\bf x},{\bf x}_0) \cdot {\bf n}({\bf x})dS({\bf x}),  
\end{equation}
\end{widetext}
where ${\bf G}({\bf x},{\bf x}_0)$ and ${\bf Q}({\bf x},{\bf x}_0)$ are free space Green's functions for Stokes equations~\cite{rallison78,poz92}. 
\begin{equation}
 {\bf G}({\bf x},{\bf x}_0)=\frac{{\bf I}}{| \hat{{\bf x}}|}+\frac{\hat{\bf x}\hat{\bf x}}{{| \hat{{\bf x}}|}^3}, \quad {\bf Q}({\bf x},{\bf x}_0) = -6\frac{\hat{\bf x}\hat{\bf x}\hat{\bf x}}{{|\hat{{\bf x}}|}^5},
 \end{equation}
are respectively termed as stokeslet and stresslet. The resultant non-hydrodynamic traction acting at the interface, $\Delta{\bf f}=-({\bf f}^{el}+Ca\ {\bf f}^E+\ {\bf f}^b$), is responsible for the deformation of the capsule.   

From the known interfacial velocity, ${\bf u}({\bf x})$ at $t$, the deformation of the capsule to a new shape at $t+\Delta t$ can be calculated through the kinematic condition 
\begin{equation}
 {\bf x}(t+\Delta t)={\bf x}(t)+k_f {\bf u}({\bf x})\Delta t, 
\end{equation}
where $k_f$ is the kinematic factor and $\Delta t$ is the step size of time used in boundary integral calculations. The kinematic factor depends upon the scaling of the variables.  

 \subsection{Analytical theory}
A standard asymptotic expansion in the small parameter, the capillary number, is carried out assuming the deformations to be proportional to the capillary number~\citep{rt16}, the details of which are provided in~\cref{sec:antheory}.

\section{Results}

\begin{table}[h]
\small
  \caption{\ Associated time scales and their reciprocals at different conductivity ratios for $C_m=50$ and $\epsilon_r=1$}
  \label{tab:times}
  \begin{tabular*}{0.48\textwidth}{@{\extracolsep{\fill}}llllll}
    \hline
    $\sigma_r$ & $t_{E,i}$	& $t_{MW}$ & $t_{cap}$  & $t_{MW}^{-1}$ & $t_{cap}^{-1}$ \\
    \hline
       0.1	&10	&1.43	&525	&0.7	&0.0019	\\
       0.3	&3.33	&1.30	&191.66	&0.77	&0.0052\\
       0.4	&2.5	&1.25	&150	&0.8	&0.0067\\
       1	&1.0	&1.0	&75	&1.0	&0.0133	\\
       10	&0.1	&0.25	&30	&4.0	&0.0333 \\
    \hline
  \end{tabular*}
\end{table}
The typical time scales associated with this problem are charge relaxation time of the external fluid medium ($\tilde{t}_E=\epsilon_e\epsilon_0/\sigma_e$), hydrodynamic response time ($\tilde{t}_H=\mu_e a/E_s$), capacitor charging time ($\tilde{t}_{cap}=a C_m[1/\sigma_i+1/2\sigma_e]$) and the Maxwell-Wagner relaxation time ($\tilde{t}_{MW}=\epsilon_0[\epsilon_i+2\epsilon_e]/[\sigma_i+2\sigma_e]$). We scale time by $\tilde{t}_E$ and  restrict our calculations to the hydrodynamic timescale equal to charge relaxation time of the external fluid medium i.e $\tilde{t}_H=\tilde{t}_E$. Therefore, the nondimensional time scales are $t_E=\tilde{t}_E/\tilde{t}_E=1$, $t_H=\tilde{t}_H/\tilde{t}_E=1$, $t_{cap}=\tilde{t}_{cap}/\tilde{t}_E=\hat{C}_m(1/2+1/\sigma_r)$ and $t_{MW}=\tilde{t}_{MW}/\tilde{t}_E=(2+\epsilon_r)/(2+\sigma_r)$. The charge relaxation time of the internal fluid is $t_{E,i}=t_E(\epsilon_r/\sigma_r)$. Unless specified, the non-dimensional membrane properties are $\hat{C}_m=50$, $\hat{G}_m=0$, and the ratio of dielectric constants $\epsilon_r=1$.  At different conductivity ratios, all the associated time scales are reported in~\cref{tab:times}. Due to the selection of scaling parameters, the kinematic factor is the inverse of the hydrodynamic timescale, i.e., $k_f=1/t_H$. In this analysis, $t_H$ is considered to be unity such that the hydrodynamic response is instantaneous to the changes in electric field.
\begin{figure}
  \includegraphics[width=0.45\textwidth]{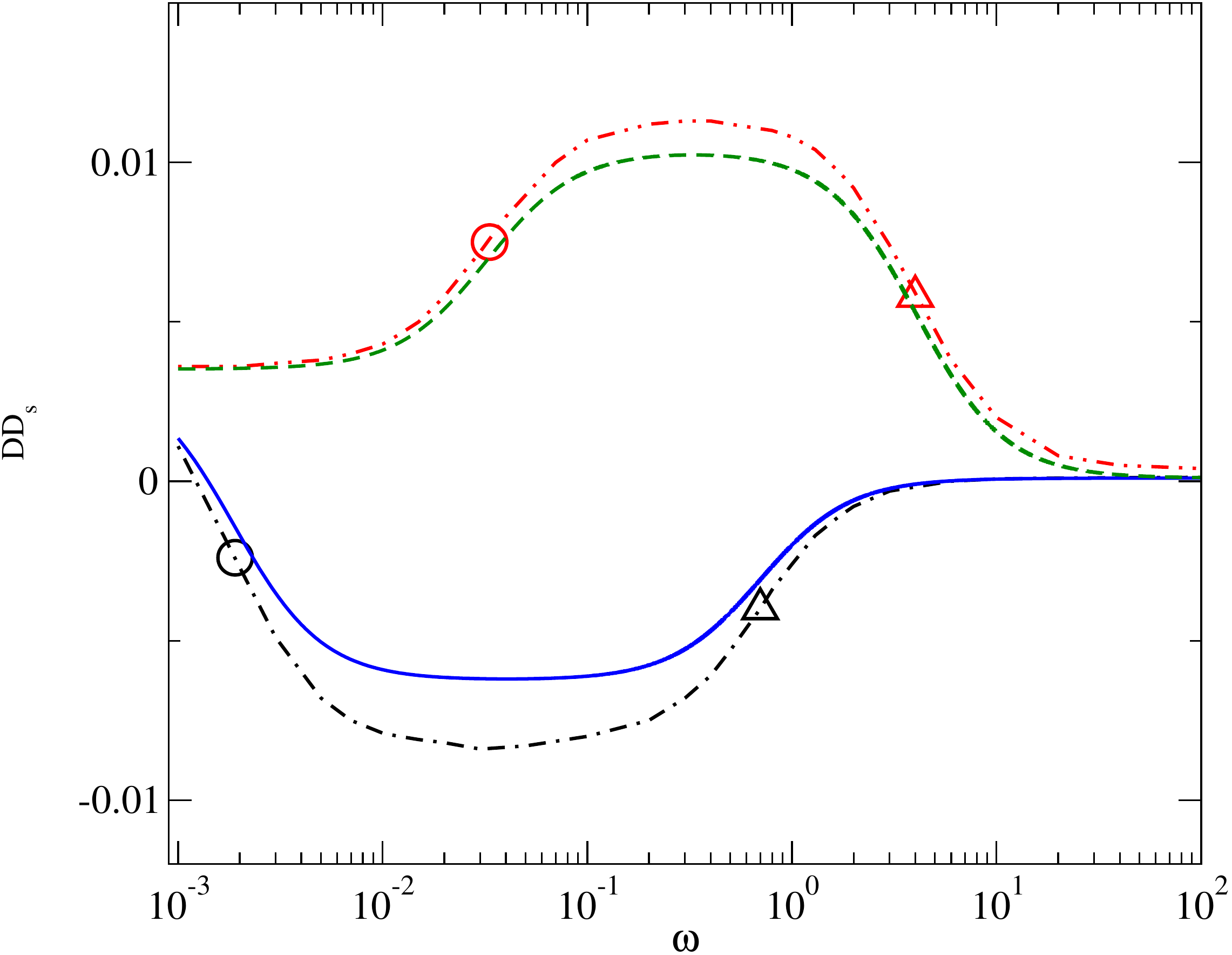}%
  \caption{Time-averaged degree of deformation at $Ca=0.01$ as a function of frequency considering $\epsilon_r=1$, $G_m=0$ and $C_m=50$. For $\sigma_r=0.1$, boundary integral and analytical results are shown by (\textcolor{black}{$\pmb{-\cdot -}$}), and (\textcolor{blue}{$\pmb{\mi}$}) curves, respectively and for $\sigma_r=10$, boundary integral and analytical results are shown by (\textcolor{red}{$\pmb{\cdot\cdot-}$}) and (\textcolor{forestgreen}{$\pmb{--}$}) curves, respectively. Marker points $\bigcirc$ represent $t_{cap}^{-1}$ and $\bigtriangleup$ represent $t_{MW}^{-1}$ for the corresponding curves.}
  \label{fig:validation}
\end{figure}

\begin{figure}
\centering
  \includegraphics[width=0.45\textwidth]{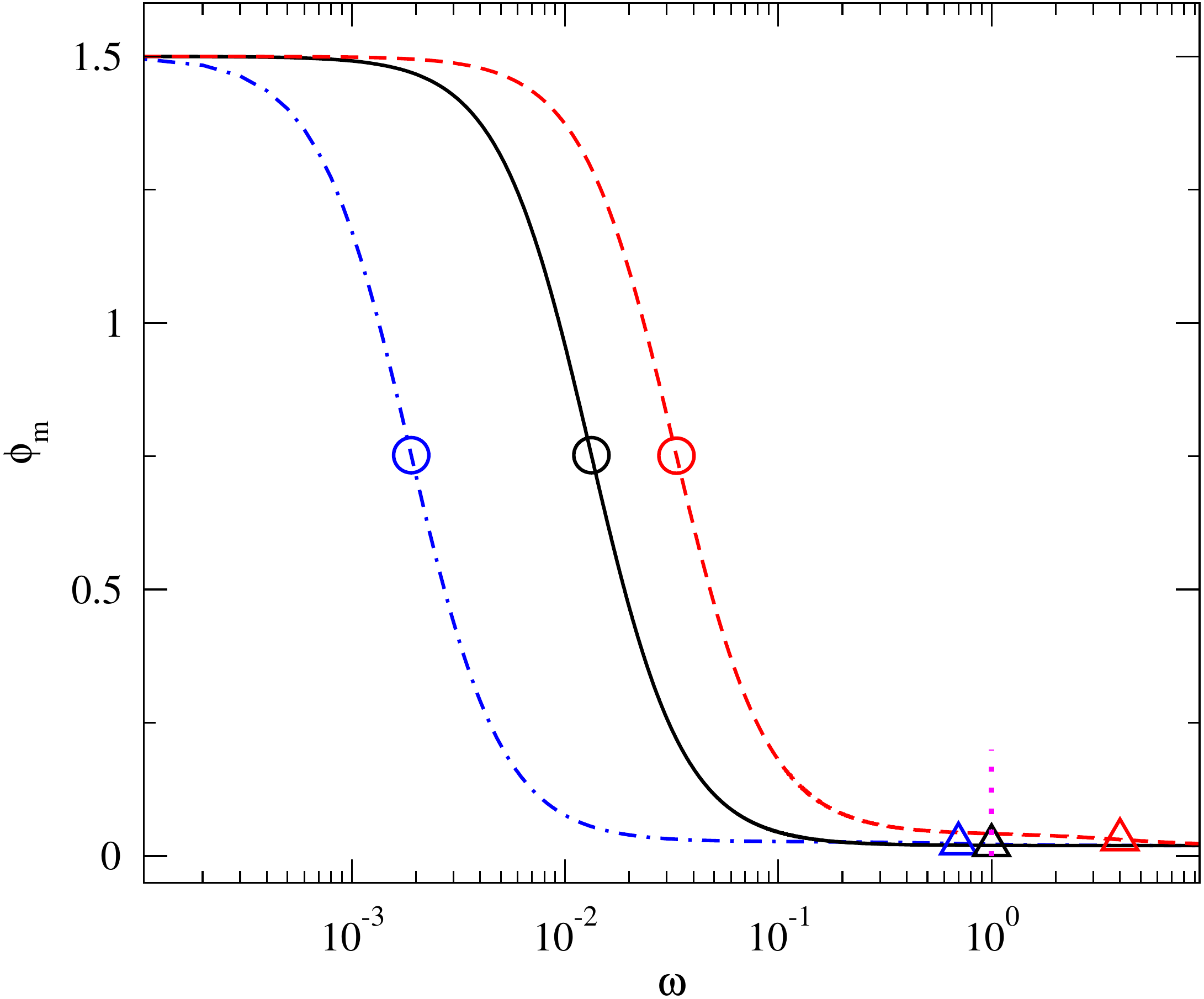}%
  \caption{Maximum value of transmembrane potential as a function of frequency considering $\epsilon_r=1$, $G_m=0$ and $C_m=50$ for $\sigma_r=0.1$ (\textcolor{blue}{$\pmb{-\cdot -}$}), $\sigma_r=1$ (\textcolor{black}{$\pmb{\mi}$}) and $\sigma_r=10$ (\textcolor{red}{$\pmb{--}$}). Marker points $\bigcirc$ represent $t_{cap}^{-1}$ and $\bigtriangleup$ represent $t_{MW}^{-1}$ for the corresponding curves. Vertical line (\textcolor{magenta}{$\pmb{\cdots}$}) at $\omega=1$ represents $t_E^{-1}$.}
  \label{fig:vmvsw}
\end{figure}

\Cref{fig:vmvsw} shows the variation of the maximum transmembrane potential ($\phi_m$ at the poles) with frequency for the case of $\sigma_r=10,1,$ and $0.1$. The expression for the $\phi_m$ is provided in~\cref{eq:phimexp} which reduces to Schwan equation~\citep{Schwan1983} when $t_E=\epsilon_e\epsilon_0/\sigma_e$ and $t_{E,i}=\epsilon_i\epsilon_0/\sigma_e\ll t_{cap}$. From the figure, it can be observed that at very low frequency, $\phi_m$ is independent of the conductivity ratio and is equal to that of a capsule in DC electric field, which is due to complete charging of a membrane at low frequencies. At very high frequencies, a membrane remains completely uncharged, and thereby $\phi_m$ attains a small value given by 
\begin{equation}
 \phi_{m,r}=\frac{3\epsilon_r}{2\hat{C}_m+\epsilon_r(2+\hat{C}_m)}.
\end{equation}
The partial charging of a membrane at the intermediate frequencies leads to frequency dependent $\phi_m$ which increases with $\sigma_r$. Therefore, a strong  dependence of the deformation on the conductivity ratio is expected.

\begin{figure}
  \includegraphics[width=0.45\textwidth]{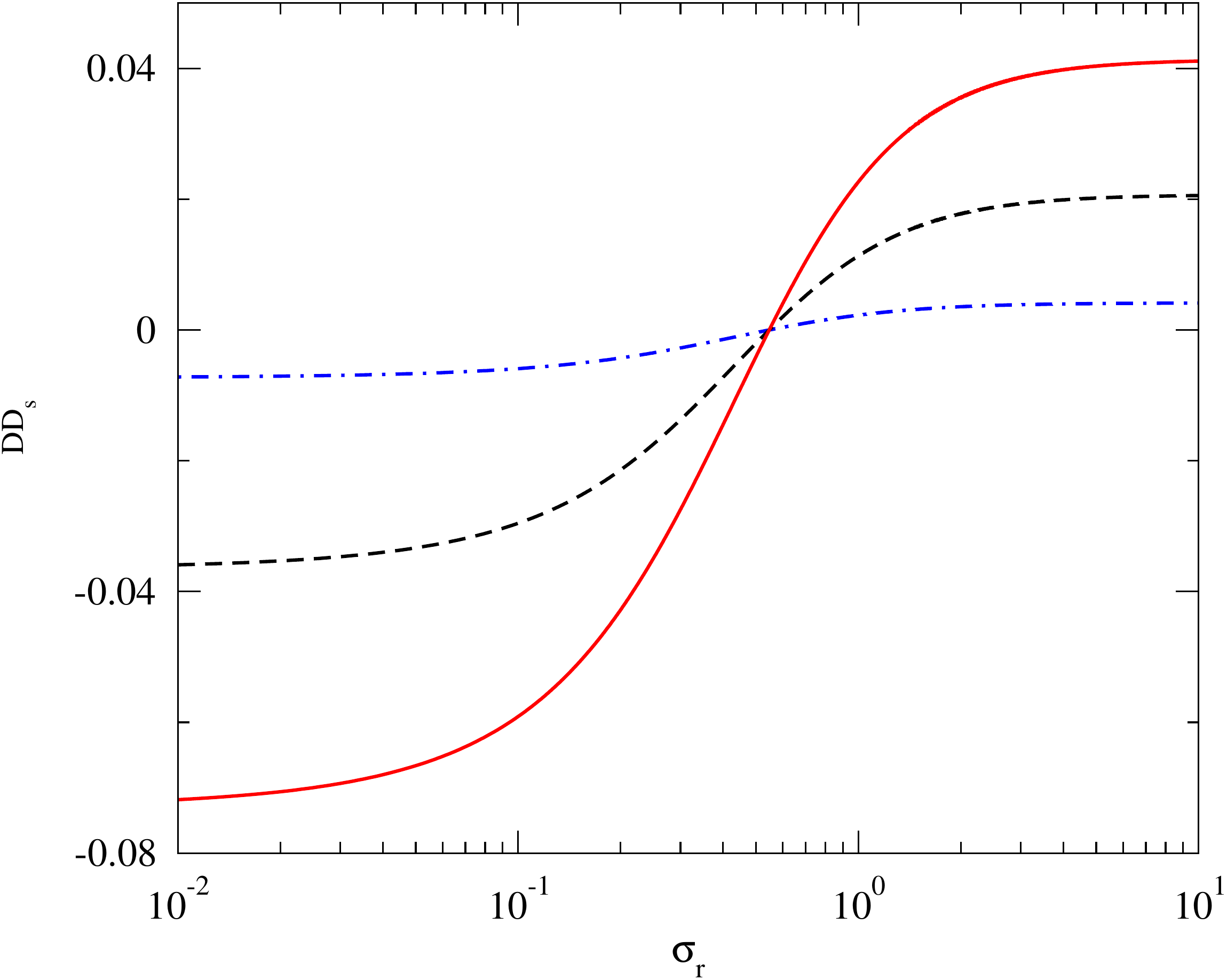}%
  \caption{Time-averaged degree of deformation of a capsule as a function of conductivity ratio at $Ca=0.01$ (\textcolor{blue}{$\pmb{-\cdot-}$}), $Ca=0.05$ (\textcolor{black}{$\pmb{--}$}) and $Ca=0.1$ (\textcolor{red}{$\pmb{\mi}$}) considering $\omega=0.1$, $\epsilon_r=1$, $C_m=50$ and $G_m=0$.}
  \label{fig:ddvssr}
\end{figure}

As a measure of deformation of the capsule, the Taylor deformation parameter is used, which is defined as the degree of deformation $DD=\frac{L-B}{L+B}$, where $L$ and $B$ are the diameters of the capsule along the rotational axis of symmetry and at the equator, respectively.  The degree of deformation can be decomposed into a stationary (time-averaged) and a time-periodic part $DD=DD_s+DD_{t} \cos{2 \omega t}$, where the frequency doubling of the deformation is a result of the square dependence of the Maxwell stress on the electric field.  The time-averaged degree of deformation as a function of the frequency of the applied electric field, obtained from boundary integral simulation and analytical theory, is compared in~\cref{fig:validation}, for $\sigma_r=10$ and $0.1$ at a low capillary number, $Ca=0.01$. 

At very high frequencies, $\omega\gg t_{E}^{-1}, t_{MW}^{-1},t_{cap}^{-1}$, the electrical impedance of the membrane is negligible, both the fluids behave as dielectrics, and the transmembrane potential is negligible. Therefore, the deformation is almost negligible on account of no dielectric contrast between the inner and the outer fluids, and a spherical geometry is observed.

\begin{figure}
  \includegraphics[width=0.45\textwidth]{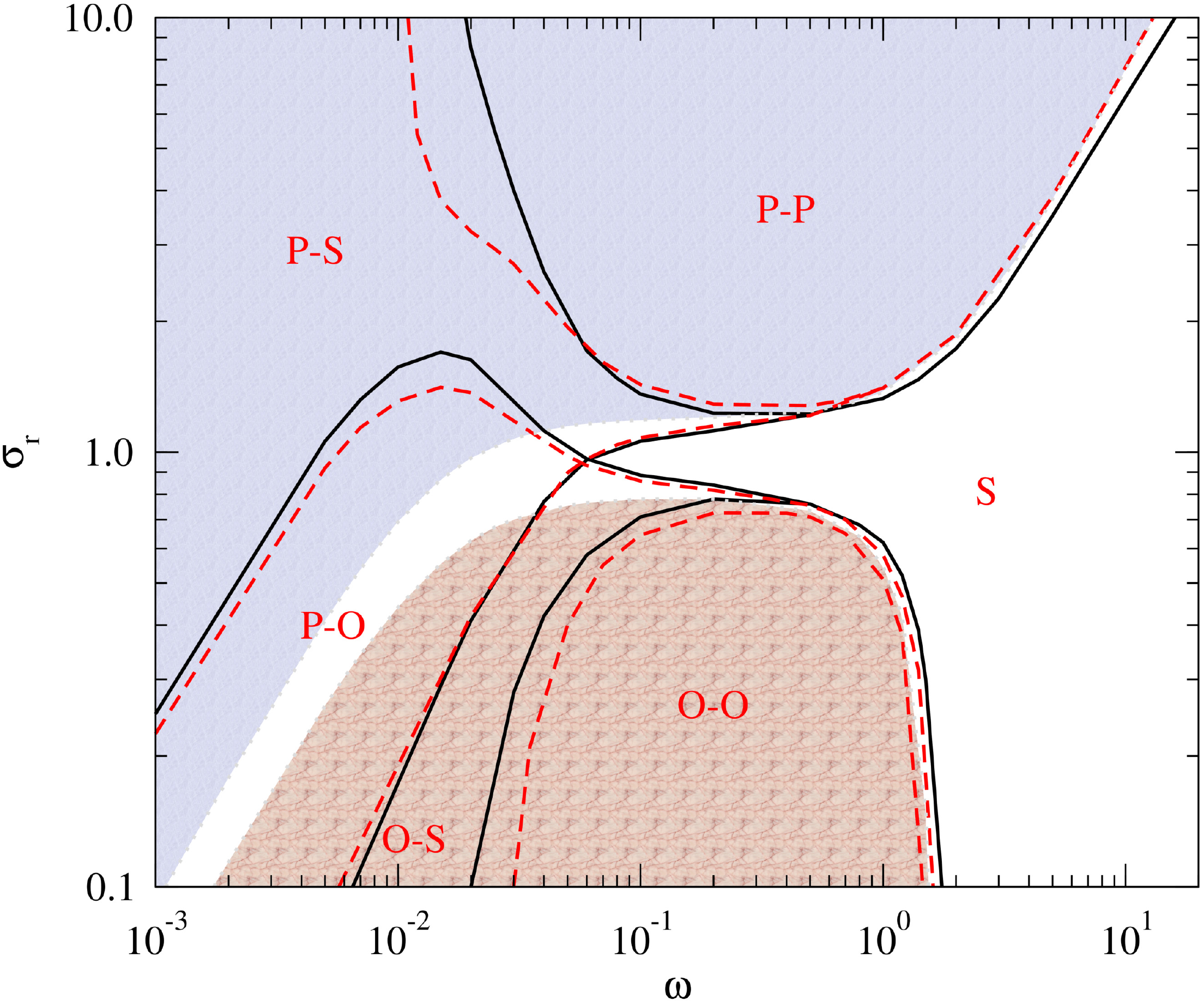}%
  \caption{Phase diagram of time-periodic breathing of a capsule at $Ca=0.1$ considering $\epsilon_r=1$, $G_m=0$ and $C_m=50$ representing prolate-prolate breathing (P-P), prolate-sphere breathing (P-S), prolate-oblate breathing (P-O), oblate-sphere breathing (O-S), oblate-oblate breathing (O-O), and undeformed sphere (S) zones of the deformation. Continuous curves are numerically obtained boundaries whereas dashed curves are analytically obtained boundaries separating zones of different modes of breathing. Shaded area towards the lower part of the plot represents time-averaged oblate deformation and the shaded area towards the upper part of the plot represents time-averaged prolate deformation.}
  \label{fig:phase0p1}
\end{figure}

 \begin{figure*}
 \begin{center}
  \includegraphics[width=1\textwidth, trim=0.0in 0.0in 0.0in 0.0in, clip]{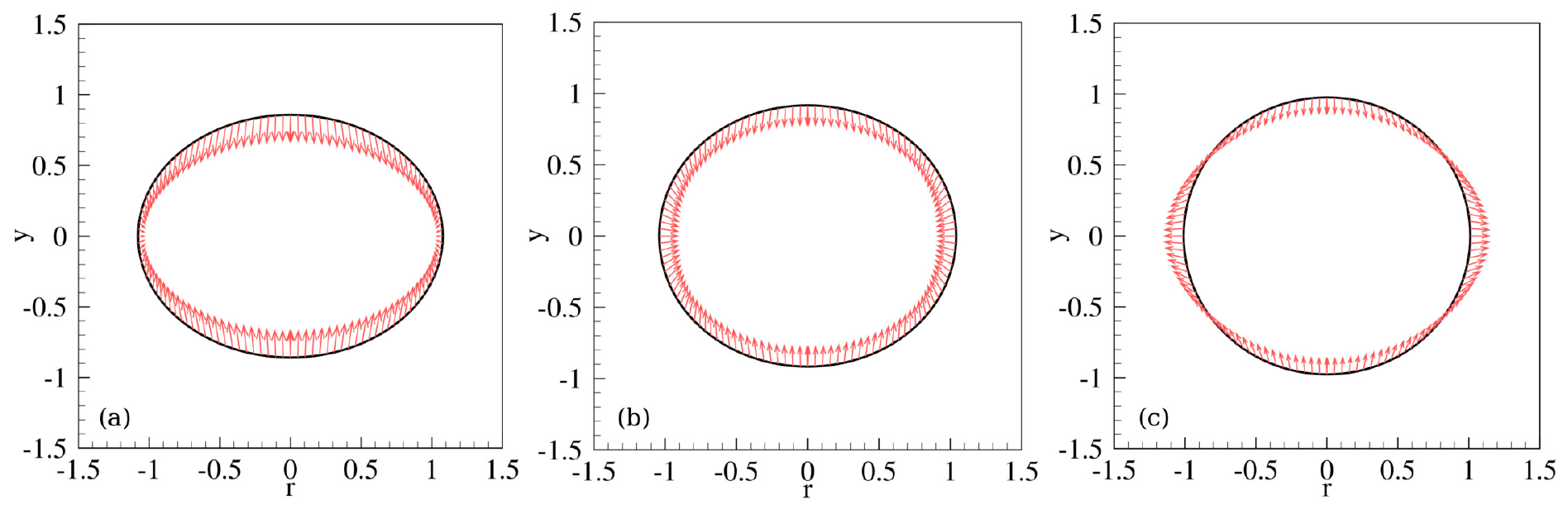}
  \caption{In the limit of small capillary number ($Ca=0.1$) for $\omega=0.04$ and $\sigma_r=0.3$ (corresponding to the C-O breathing at high capillary number) considering $\epsilon_r=1$, $G_m=0$ and $C_m=50$, electric stresses (arrows at the interface) are shown for the shapes observed during O-O breathing at $t=477$, $496$ and $515$.}
\label{fig:strsqo0p1}
\end{center}
\end{figure*} 

At frequencies $\omega>t_{cap}^{-1}$, the impedance of the membrane is still very low, and the charging of the membrane is governed by the electrical properties of the inner and outer fluid media. The  fluids show a leaky dielectric behavior in the bulk, such that the outer fluid exhibits conductive behavior at frequencies lower than $t_E^{-1}=1$, whereas the inner at frequencies lower than $t_{E,i}^{-1}$. As the frequency is lowered, the impedance of the membrane increases and  a net  interfacial charge build-up starts at a frequency $t_{MW}^{-1}$. The interface at frequency $t_{MW}^{-1}$ is similar to that of a liquid drop on the timescale of $t_{MW}$ and admits a net free charge and tangential stress. Depending upon the conductivity ratio, the tangential stress can be from poles to the equator ($\sigma_r<1$) or equator to poles ($\sigma_r>1$), thereby imparting an oblate spheroid or a prolate spheroid shapes, respectively~(\cref{fig:validation}). This regime also corresponds to the maximum deformation of the capsule. 

As the frequency is further decreased, $\omega<t_{cap}^{-1}$, the membrane now becomes charged, the membrane  impedance is fairly high such that the normal electric fields inside  and outside the capsule tend to disappear, on account of the insulating membrane. The tangential stresses, therefore, become negligible and the capsule behaves like a dielectric drop in a conducting fluid. This drop in DC field-like behavior implies that the conductivity ratio does not play a role anymore and prolate spheroids are obtained. A fairly good agreement is obtained between analytical theory (valid at small capillary numbers) and boundary integral calculations.

To verify whether the boundary of the time-averaged prolate to oblate deformation transition is affected by the capillary number, $DD_s$ vs. $\sigma_r$  plot is shown for different capillary numbers at a given frequency~(\cref{fig:ddvssr}). For a particular frequency (in this case $\omega=0.1$), the time-averaged deformation changes from oblate to prolate with an increase in the conductivity ratio. According to the eq.~\ref{eq:ddf}, the time-averaged degree of deformation (obtained from analytical theory at small deformation) is proportional to the capillary number and the numerator is only a function of $\sigma_r$ for the specified $\epsilon_r$, $\omega$, $\hat{C}_m $, therefore, the transitional $\sigma_r$ does not change with the capillary number.

\begin{figure*}
 \begin{center}
\begin{subfigure}{.32\textwidth}
  \centering
  \includegraphics[width=1\textwidth, trim=0.0in 0.0in 0.0in 0.0in, clip]{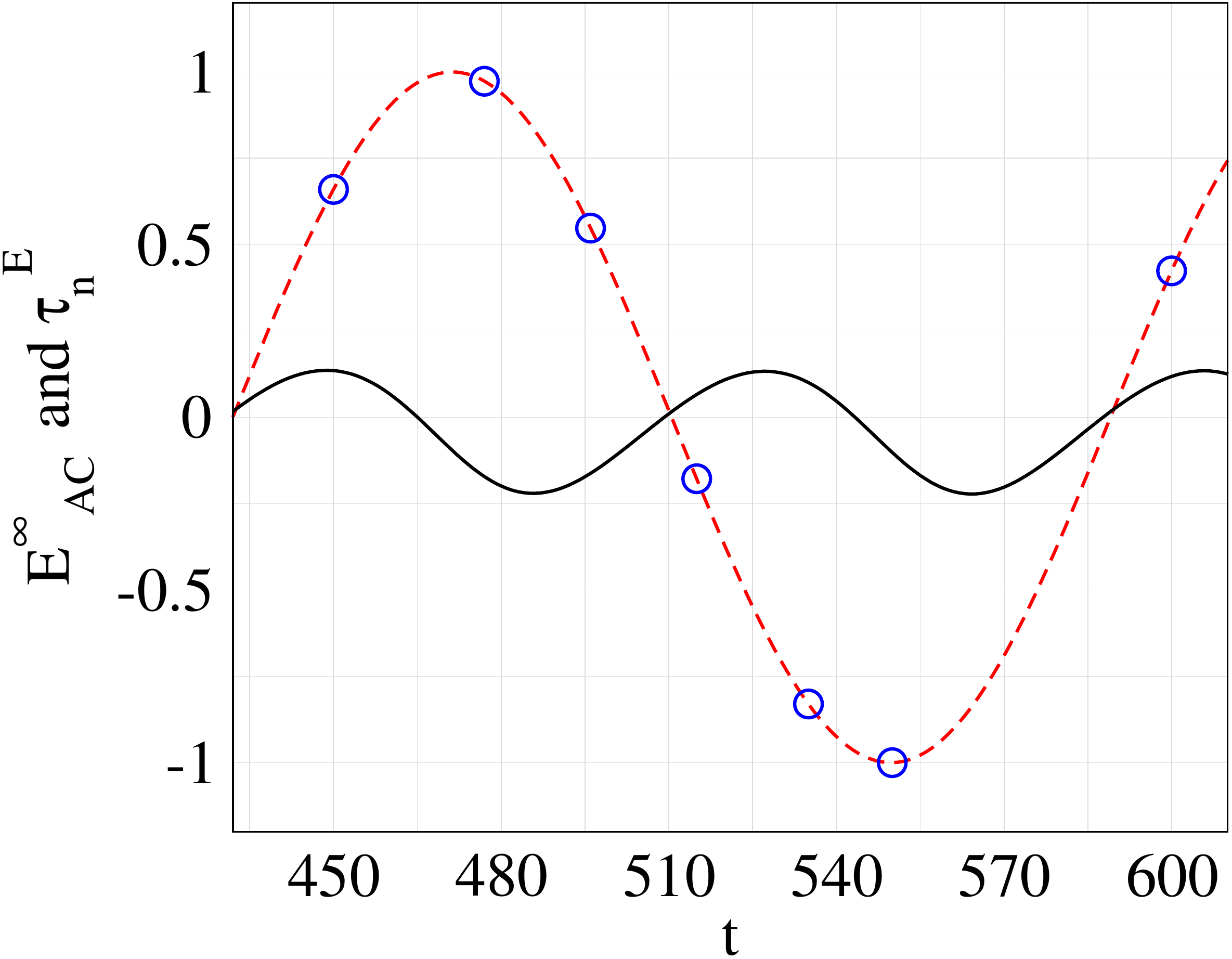}
  \caption{$\theta=\pi/2$}
  \label{fig:fieldsta}
\end{subfigure}
\begin{subfigure}{.32\textwidth}
  \centering
  \includegraphics[width=1\textwidth, trim=0.0in 0.0in 0.0in 0in, clip]{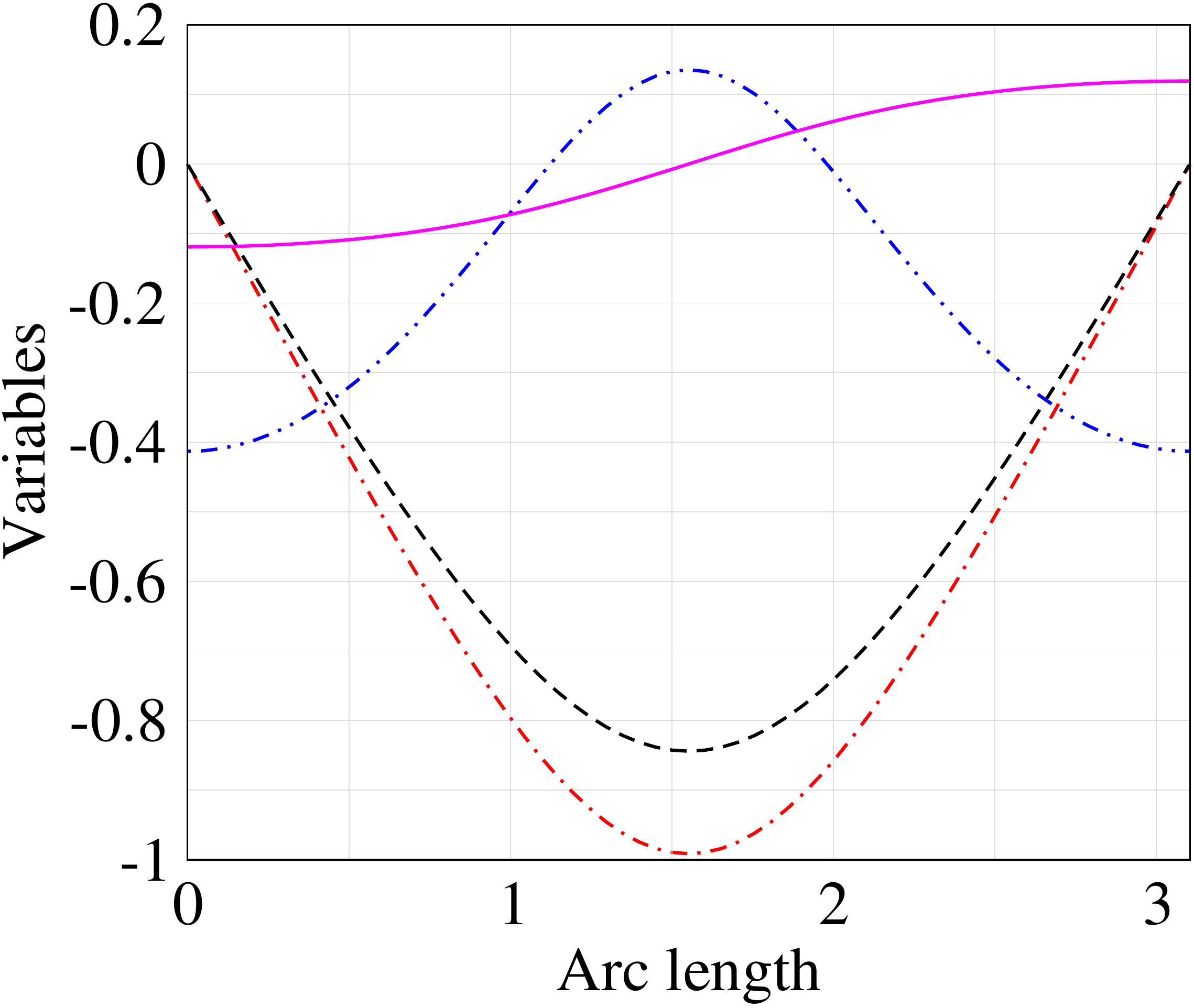}
  \caption{$t=450$}
  \label{fig:fieldstb}
\end{subfigure}
\begin{subfigure}{.32\textwidth}
  \centering
  \includegraphics[width=1\textwidth, trim=0.0in 0.0in 0.0in 0in, clip]{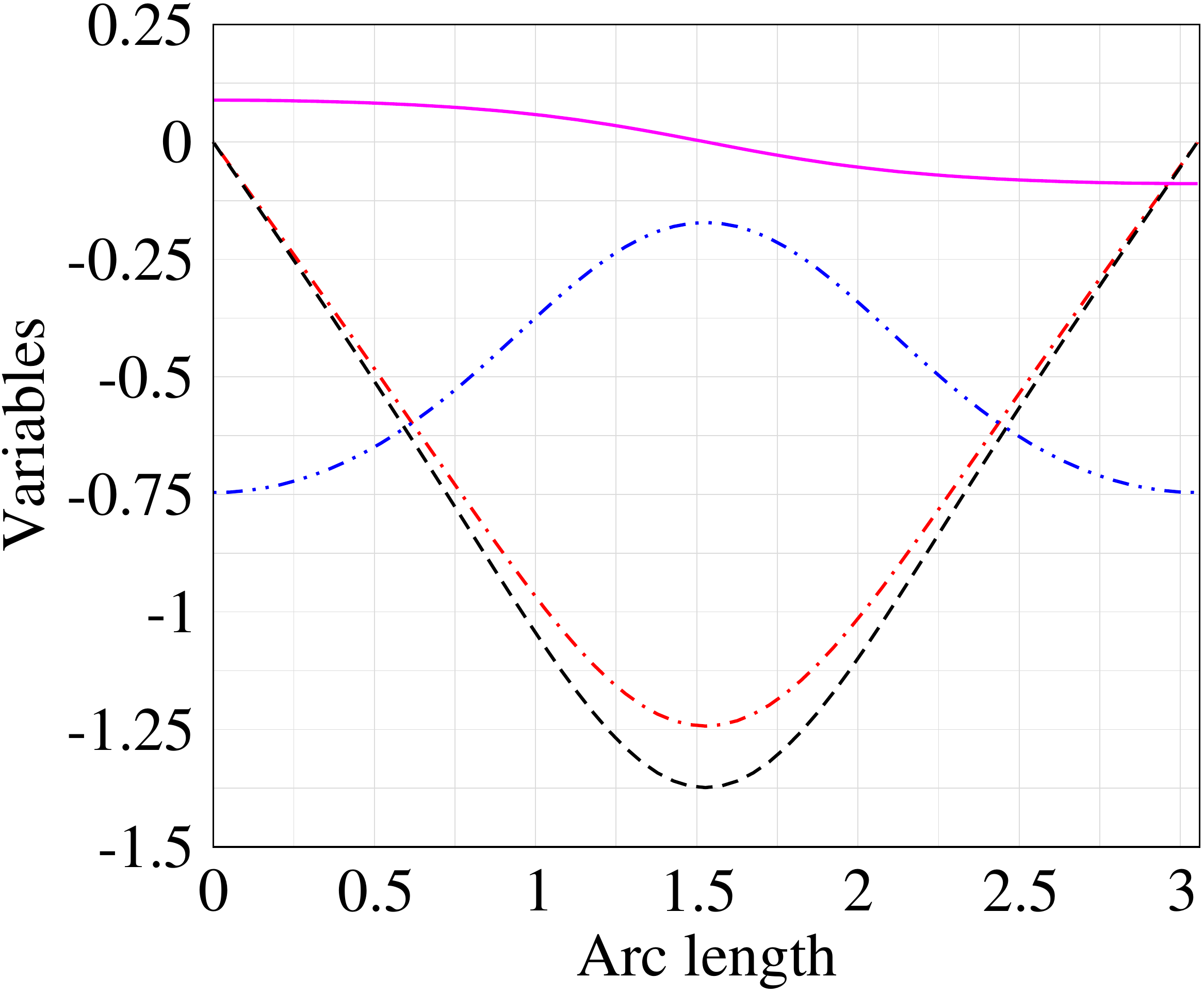}
  \caption{$t=477$}
  \label{fig:fieldstc}
\end{subfigure}
\begin{subfigure}{.32\textwidth}
  \centering
  \includegraphics[width=1\textwidth, trim=0.0in 0.0in 0.0in 0in, clip]{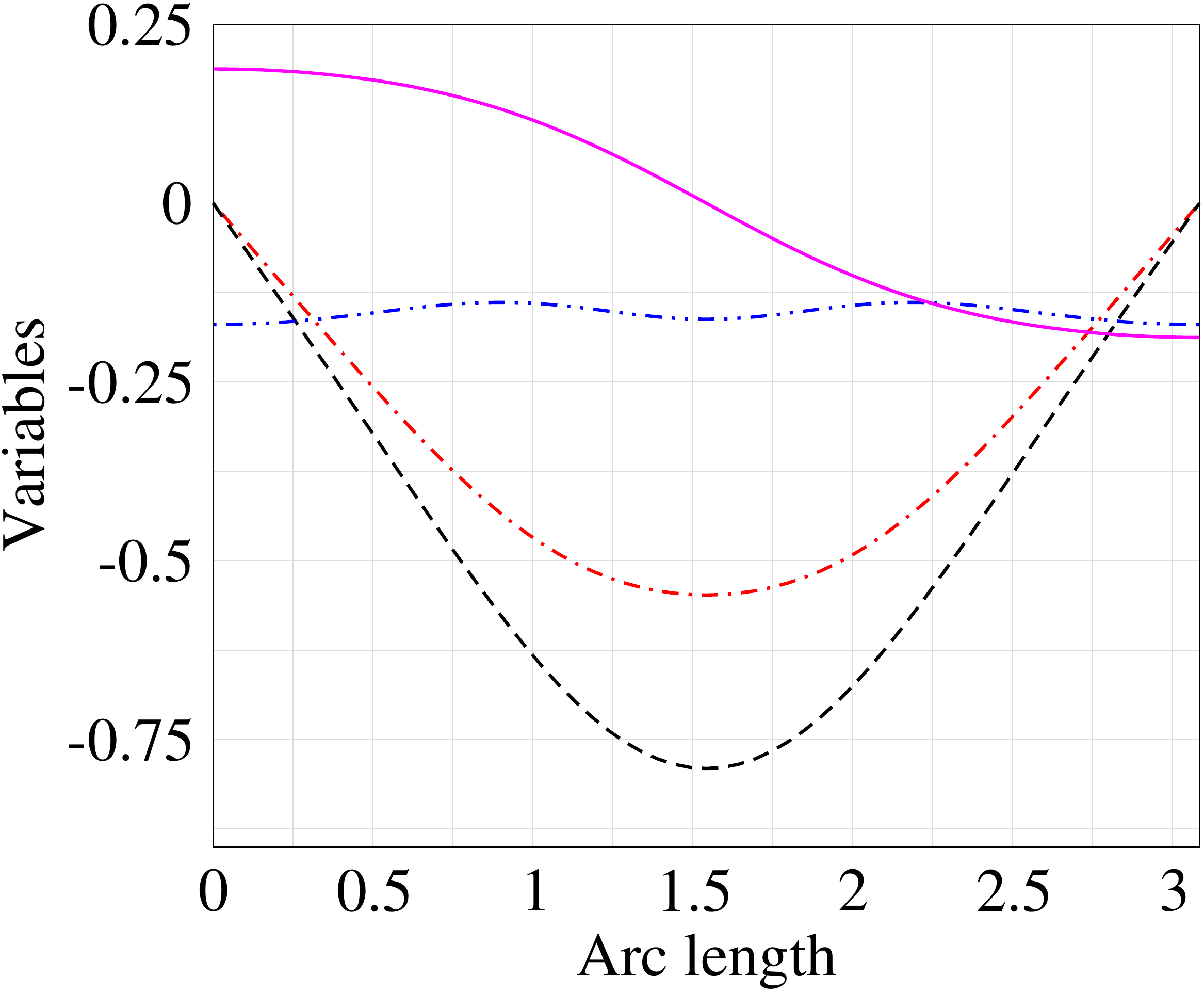}
  \caption{$t=496$}
  \label{fig:fieldstd}
\end{subfigure}
\begin{subfigure}{.32\textwidth}
  \centering
  \includegraphics[width=1\textwidth, trim=0.0in 0.0in 0.0in 0.0in, clip]{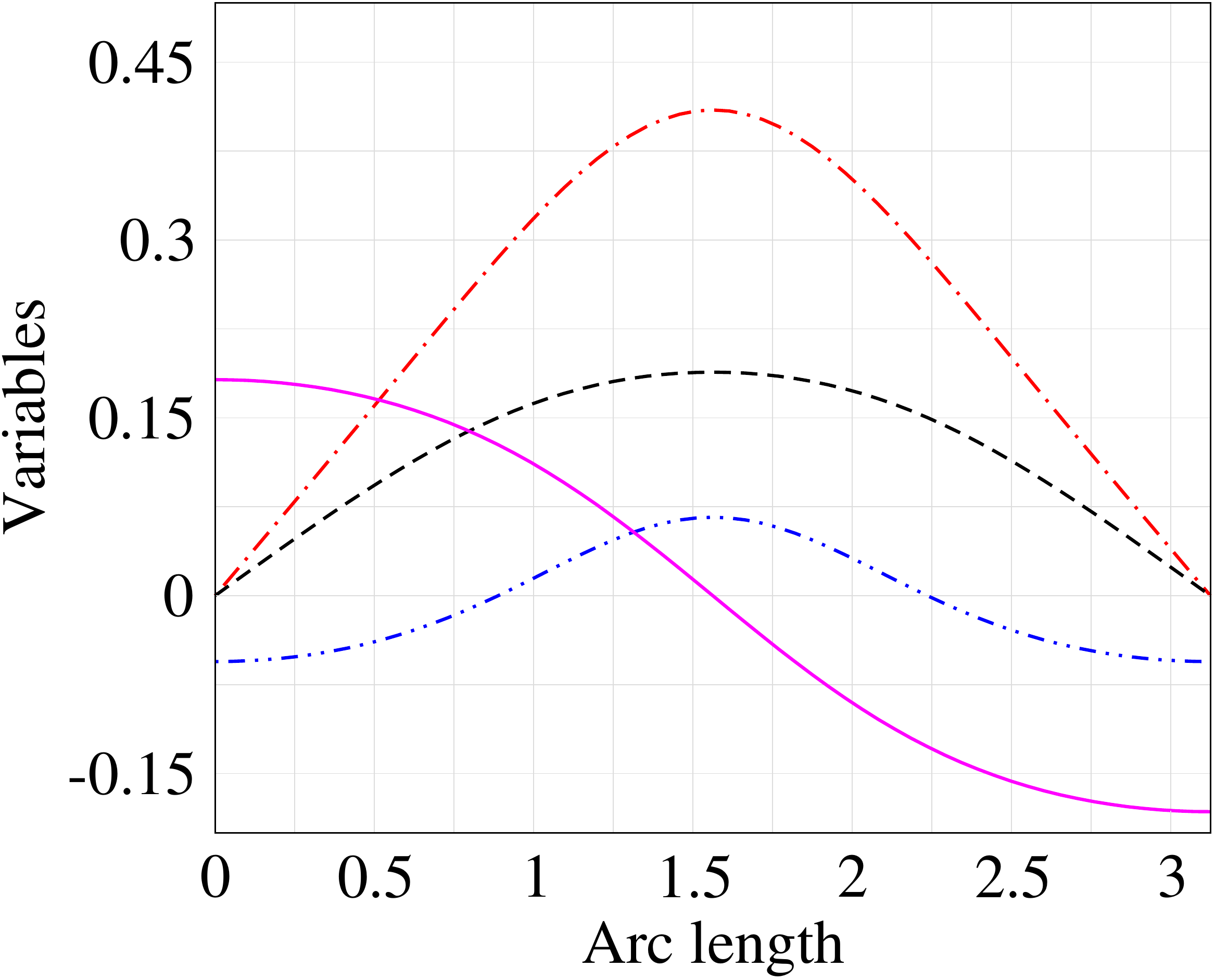}
  \caption{$t=515$}
  \label{fig:fieldste}
\end{subfigure}
\begin{subfigure}{.32\textwidth}
  \centering
  \includegraphics[width=1\textwidth, trim=0.0in 0.0in 0.0in 0in, clip]{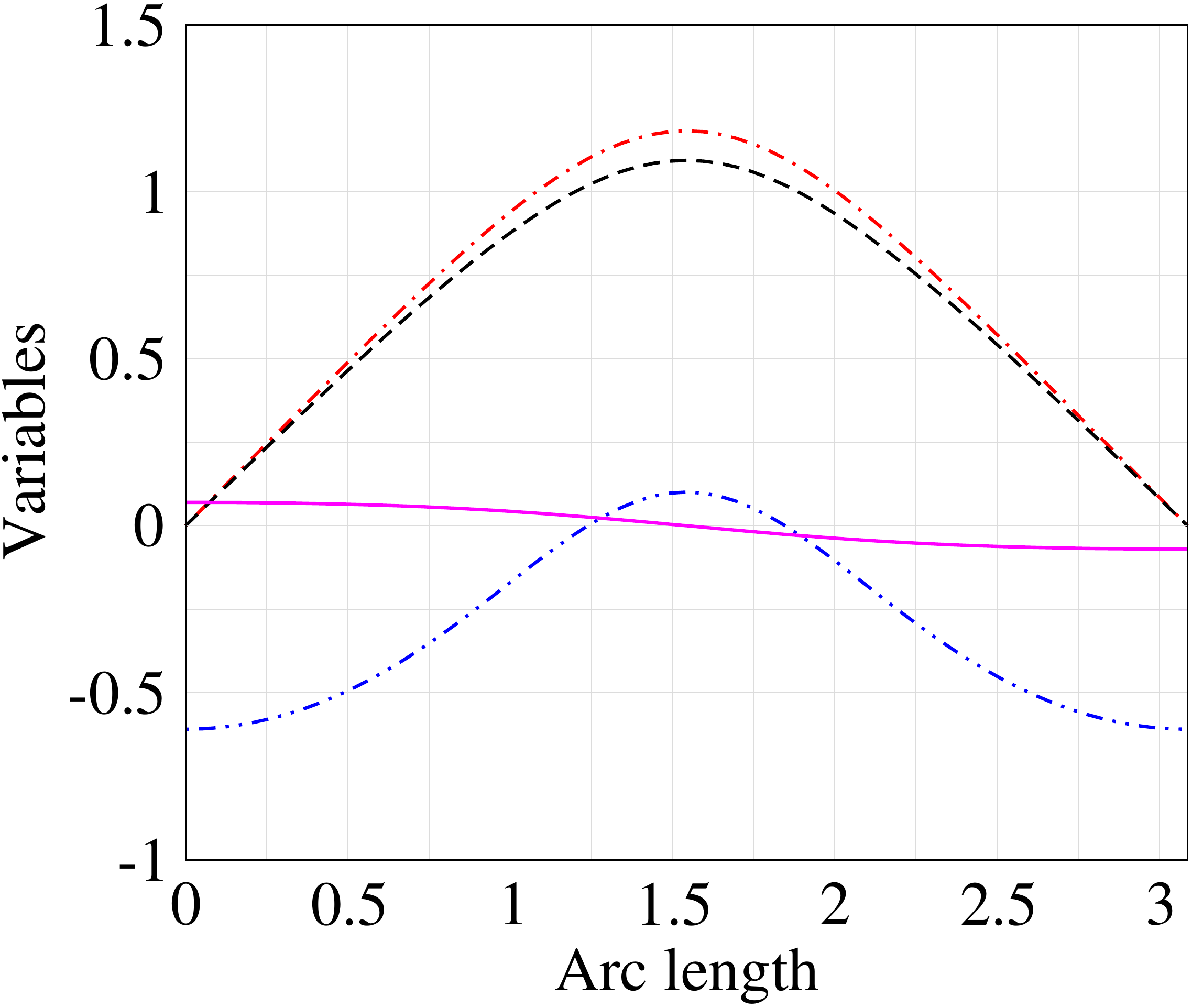}
  \caption{$t=535$}
  \label{fig:fieldstf}
\end{subfigure}
\begin{subfigure}{.32\textwidth}
  \centering
  \includegraphics[width=1\textwidth, trim=0.0in 0.0in 0.0in 0in, clip]{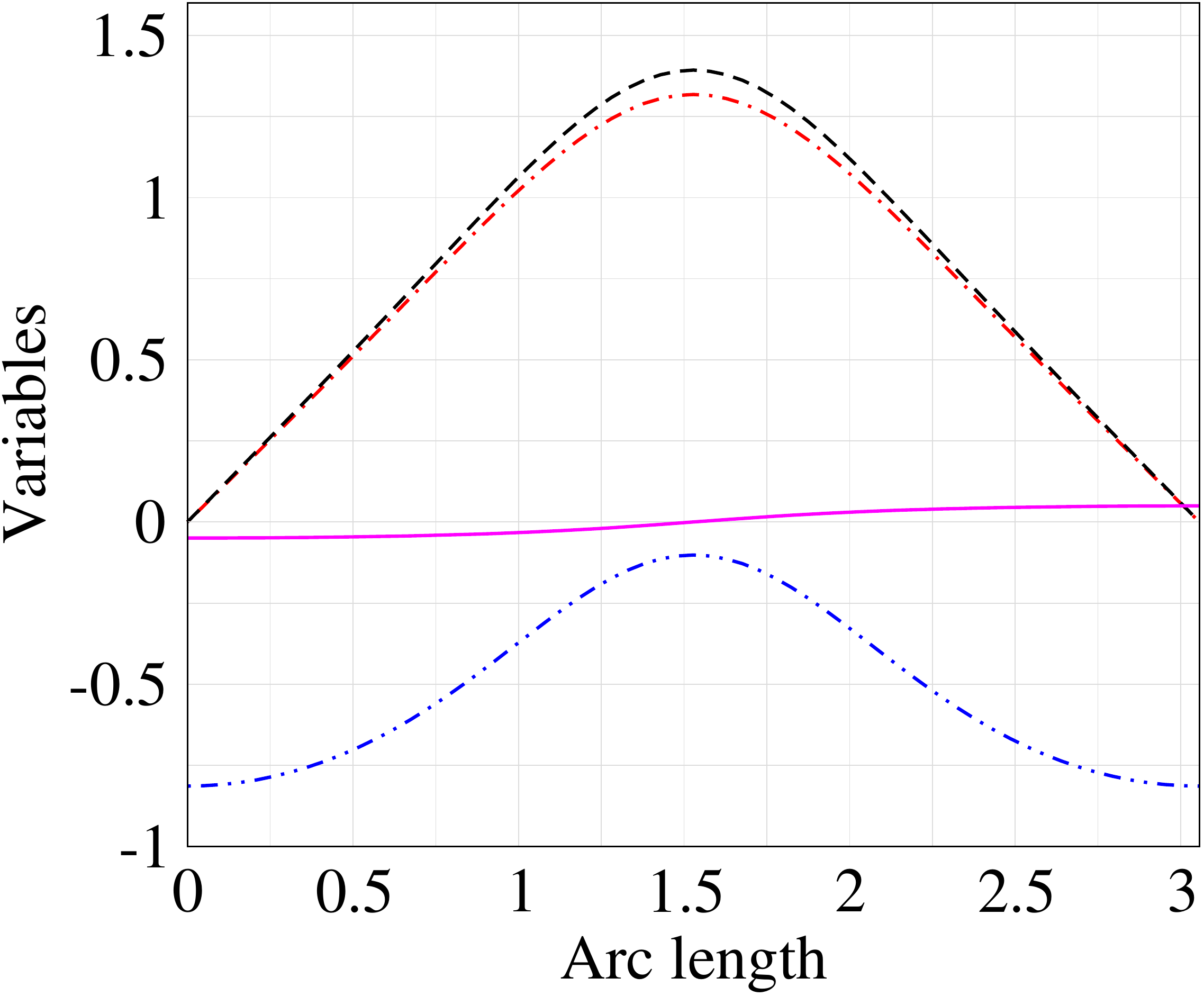}
  \caption{$t=550$}
  \label{fig:fieldstg}
\end{subfigure}
\begin{subfigure}{.32\textwidth}
  \centering
  \includegraphics[width=0.9\textwidth, trim=0.0in 0.0in 0.0in 0in, clip]{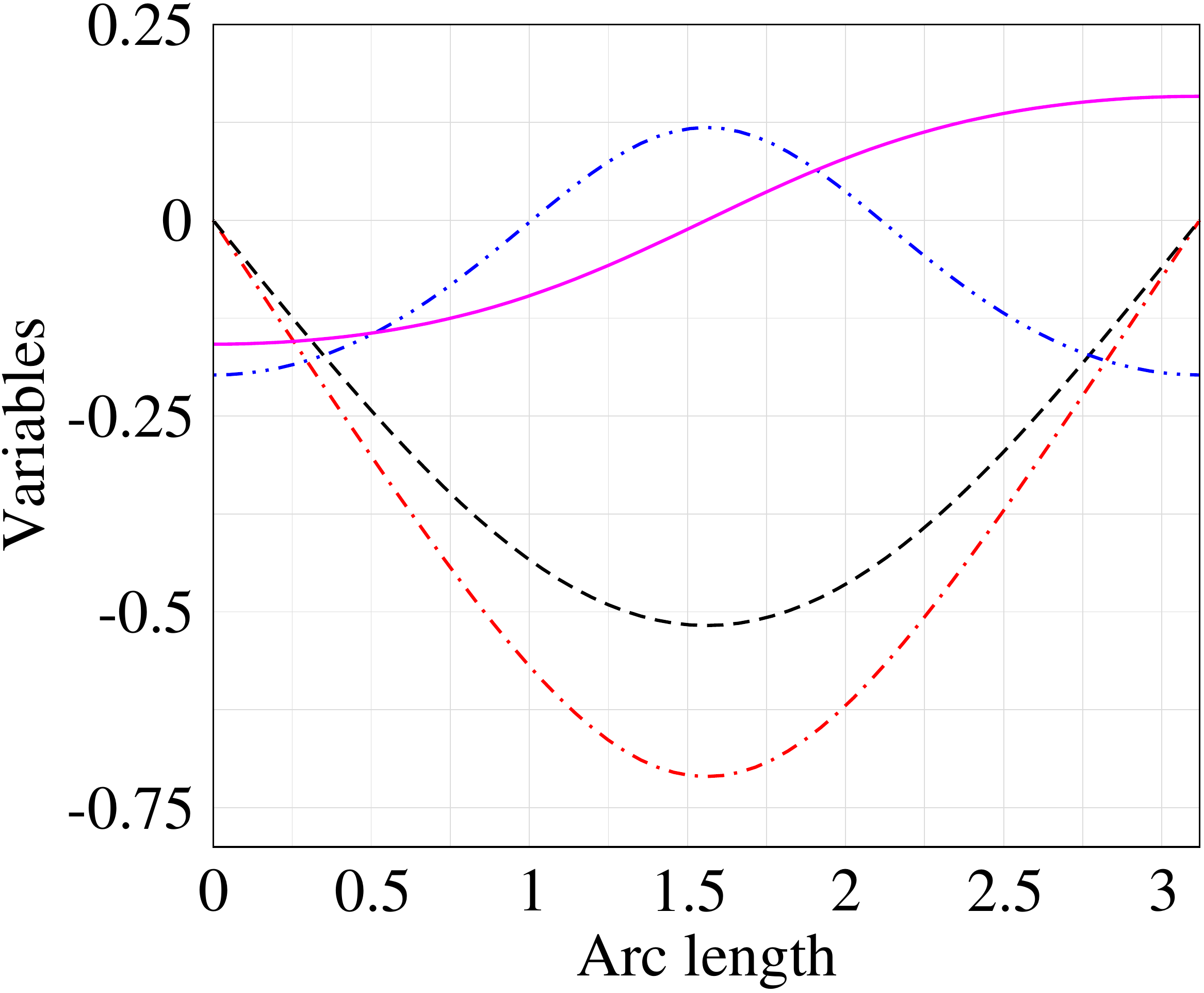}
  \caption{$t=600$}
  \label{fig:fieldsth}
\end{subfigure}
\caption{In the limit of small deformation of a capsule ($Ca=0.1$) at $\omega=0.04$ and $\sigma_r=0.3$, considering $\epsilon_r=1$, $G_m=0$ and $C_m=50$, applied electric field (\textcolor{red}{$\pmb{--}$}) and the developed normal electric stress at the equator (\textcolor{black}{$\pmb{\mi}$}) as a function of time are shown in fig. (a), marker points show the instantaneous field for the subsequent figures. In other figures the variation of transmembrane potential (\textcolor{magenta}{$\pmb{\mi}$}), inner tangential electric field (\textcolor{red}{$\pmb{-\cdot-}$}), outer tangential electric field (\textcolor{black}{$\pmb{--}$}) and normal electric stress (\textcolor{blue}{$\pmb{-\cdot\cdot-}$}) over the arc length are shown at the same conditions.}
\label{fig:fieldst}
\end{center}
\end{figure*} 

The different modes of time-averaged deformation, namely sphere, prolate and oblate spheroids can be plotted onto a phase diagram with conductivity ratio and frequency as the coordinates. \Cref{fig:phase0p1} is plotted with the boundary integral simulation results conducted for a fixed capillary number ($Ca\lesssim 0.1$) showing different regions marked out for the different modes of deformation. The different regions that are marked out by color shades in the phase diagram show different modes of time-averaged prolate and oblate deformations, respectively, whereas the white region mostly at the right half of the phase diagram represents the undeformed sphere region. The phase diagram is constructed assuming the shape to be a sphere if the degree of deformation is less than $1\%$ of the initial spherical shape. The prolate shapes in the phase diagram are obtained when $\sigma_r>1$, due to tensile normal electric stress at the poles. For $\sigma_r<1$, oblate shapes are seen at intermediate frequencies, that is $t_{cap}^{-1}<\omega<t_{MW}^{-1}$ due to compressive normal stresses at the poles. At very high frequencies $\omega>t_{MW}^{-1}$, spherical shapes are seen for all the values of $\sigma_r$. For $\sigma_r>1$, prolate shapes are seen for $\omega<t_{cap}^{-1}$, due to compressive normal electric stress at the equator. 

Unlike the deformation of a capsule in DC fields which show time-independent deformation, the AC fields show time-dependent deformation which could be termed as "breathing" between two extreme shapes. Thus the phase diagram now exhibits different  modes of breathing, namely prolate-prolate (P-P), prolate-sphere (P-S), prolate-oblate (P-O), oblate-sphere (O-S), oblate-oblate (O-O) and an undeformed sphere (S). The boundary integral simulation results (continuous curves) as well  as, the results obtained from analytical theory (dashed curves) for a fixed capillary number ($Ca=0.1$) are shown in~\cref{fig:phase0p1}. Numerically obtained phase boundaries separating zones for different breathing modes agree reasonably well with the analytically obtained results.  

The temporal shape change of a capsule during an AC cycle is essentially because of the variation of the applied electric field with time. Thus, if the hydrodynamic and electric relaxation is fast as compared to the applied frequency, one can expect the behavior of a capsule in an AC field to be  similar to the deformation of a capsule in a  DC field of equivalent strength. In this limit, when the applied field goes to zero, the deformation too should go to zero, and only the prolate-sphere or oblate-sphere modes are expected. These modes of deformation are indeed seen in~\cref{fig:phase0p1} which also shows that the P-S and O-S modes of deformation are in agreement with the time-averaged regions of the phase diagram.  However, when the hydrodynamic and electric relaxation timescales are of the same order or greater than the inverse of applied frequency, the instantaneous deformation need not be zero even when the instantaneous field is zero. This phenomena is, for example, illustrated by the P-P mode in the phase diagram~(\cref{fig:phase0p1}). Interestingly the shape can also change between two extremes, i.e., between prolate and oblate shapes and correspondingly, a variety of breathing modes can be obtained.

In general, the dynamics of a capsule in an AC field can be mapped on to the time evolution of a capsule in a DC field by realizing that the frequency in AC fields is identical to the inverse of time in DC fields. However, unique features typical of the AC fields can still be observed, which is due to the reversal of the electric field. Thus if the charge dynamics is slower than the applied frequency, a reversal of the sign of electric field (say, negative) can be observed, while the charge distribution continues to correspond to the earlier field (i.e., positive). This distributed charge causes a stress distribution over the interface and thereby, the deformation, which may not be seen in the corresponding DC fields.

Figures~\ref{fig:strsqo0p1}a-c show the stress distribution at the interface of a capsule at $\sigma_r=0.3$. At the same conductivity ratio,~\cref{fig:fieldst}a represents the applied electric field and electric stress as a function of time and figs.~\ref{fig:fieldst}b-h show the change of electrical variables on the interface of the capsule at different times. The compressive normal Maxwell stress at the poles leads to an oblate deformation. The normal electric stress at the poles is compressive on account of the higher conductivity of the external fluid medium. However, when the shape relaxes due to a decrease in the instantaneous field, the charge dynamics is such that although the applied electric field changes sign, the transmembrane potential does not. Figures~\ref{fig:fieldst}b, e, f and h indicate that the $\phi_m$ does not change sign even when the direction of the field is changed, e.g., negative $\phi_m$ is seen at north pole even when the field is in the positive $y$-direction. This negative transmembrne potential leads to the tangential electric field inside being higher than that of the outside, leading to tensile normal electric stress. Such a phenomenon is not observed in DC field which does not switch direction. Although the shape remains oblate leading to the O-O breathing mode (shown in~\cref{fig:oo}), the stress varies from compressive at the poles to tensile at the equator during a voltage cycle. Thus, the deformation modes in AC fields could have a very complex physical origin. The dynamics of the shapes in the P-O mode (shown in~\cref{fig:po}) indicates that at low frequencies, the time-averaged oblate shape relaxes to a prolate shape within a voltage cycle. The stress distribution is seen to be a sensitive function of the relative values of the membrane charging time and the Maxwell Wagner charge relaxation time with respect to the applied frequency (see figs.~\ref{fig:strsqo0p1}, figs.~\ref{fig:strsqp0p1} and figs.~\ref{fig:ca0p1bic}). 

\begin{figure}
  \includegraphics[width=0.45\textwidth]{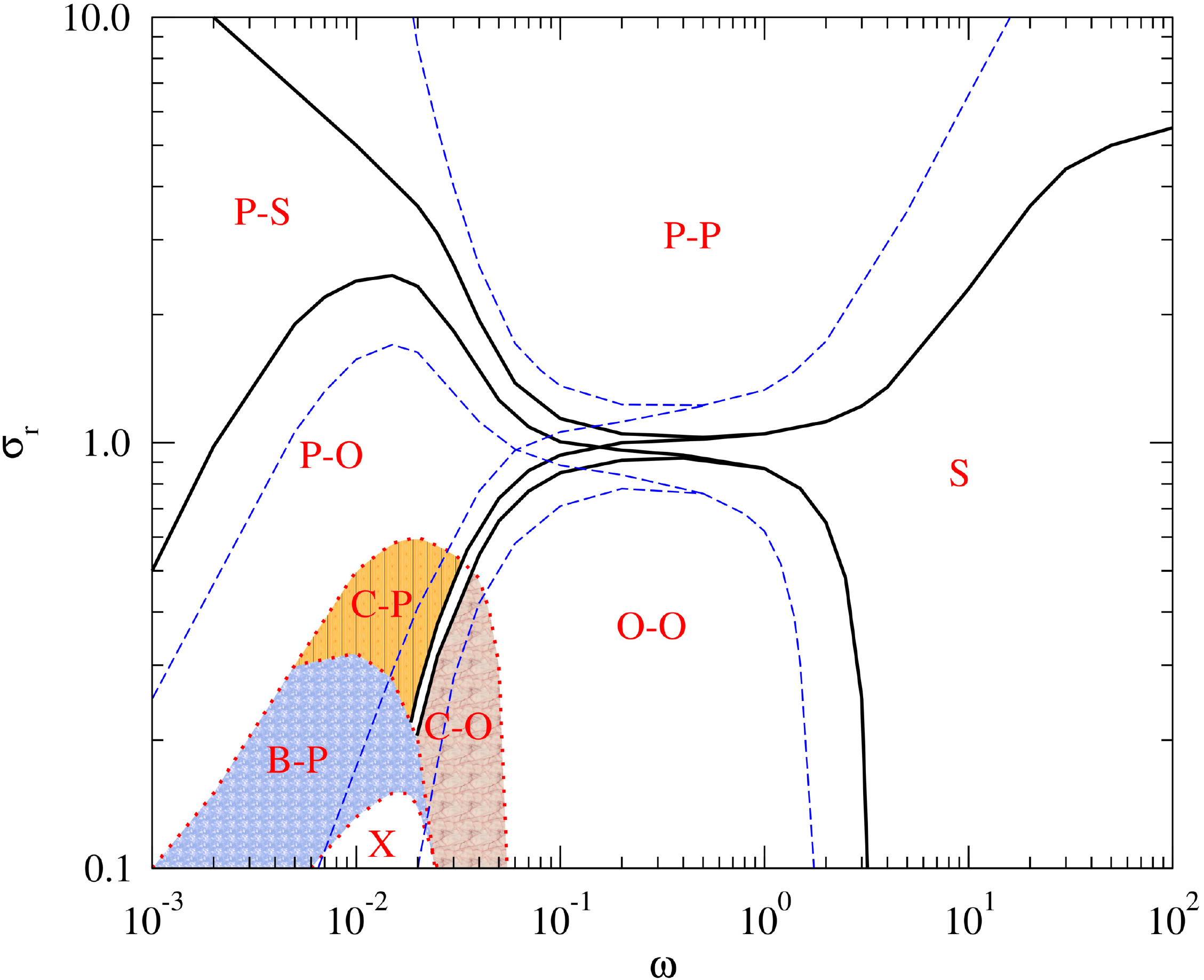}%
  \caption{Phase diagram of time-periodic breathing of capsule at $Ca=0.4$ considering $\epsilon_r=1$, $G_m=0$ and $C_m=50$ representing prolate-prolate breathing (P-P), prolate-sphere breathing (P-S), prolate-oblate breathing (P-O), oblate-oblate breathing (O-O) and undeformed sphere (S) zones of the deformation. Cylinder-prolate (C-P), cylinder-oblate (C-O), biconcave-prolate (B-P) and breakup (X) zones are shown with shades. Dashed curves are the boundaries separating zones of the phase diagram at $Ca=0.1$.}
  \label{fig:phase0p4}
\end{figure}

A similar phase diagram (only the time-periodic) of a capsule deformation is observed at the high capillary number as shown in~\cref{fig:phase0p4}. The small deformation and the large deformation phase boundaries in both the phase diagrams (\cref{fig:phase0p1} and~\cref{fig:phase0p4}) nearly overlap. 
The P-P, P-O, and O-O modes have a similar physical mechanism as in the low capillary case. The large deformations encountered in these modes are shown in figures \cref{fig:pp,fig:oo,fig:po}
in the appendix. However, a significant deviation is observed in the range of $10^{-3}\lesssim\omega\lesssim 0.5\times10^{-1}$ and $0.1\lesssim\sigma_r\lesssim 0.6$. In this range of the frequency and conductivity ratio, new modes of complex breathing are observed, namely cylinder-prolate (C-P), cylinder-oblate (C-O), and biconcave-prolate (B-P).

\begin{figure*}
  \centering
  \includegraphics[width=0.9\textwidth]{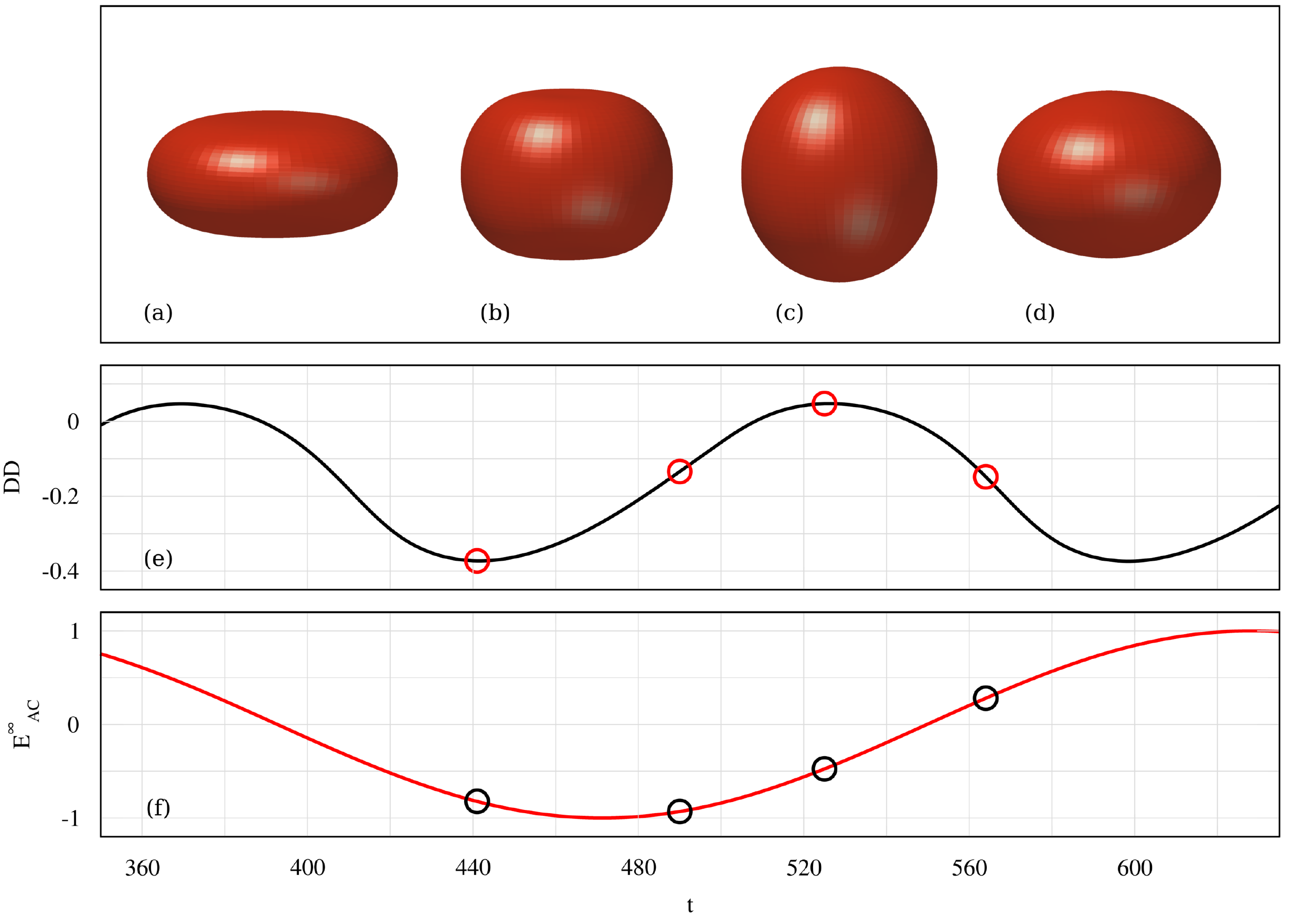}
  \caption{Cylinder-prolate (C-P) breathing of a capsule (a-d) at $Ca=0.4$ for $\omega=0.02$ and $\sigma_r=0.4$ considering $\epsilon_r=1$, $G_m=0$ and $C_m=50$. Shapes are corresponding to the marker points on the curves in (e) representing degree of deformation and in (f) representing applied electric field as the functions of time.  }
  \label{fig:cp_shape}
 \end{figure*}
 
  \begin{figure*}
  \centering
  \includegraphics[width=0.9\textwidth]{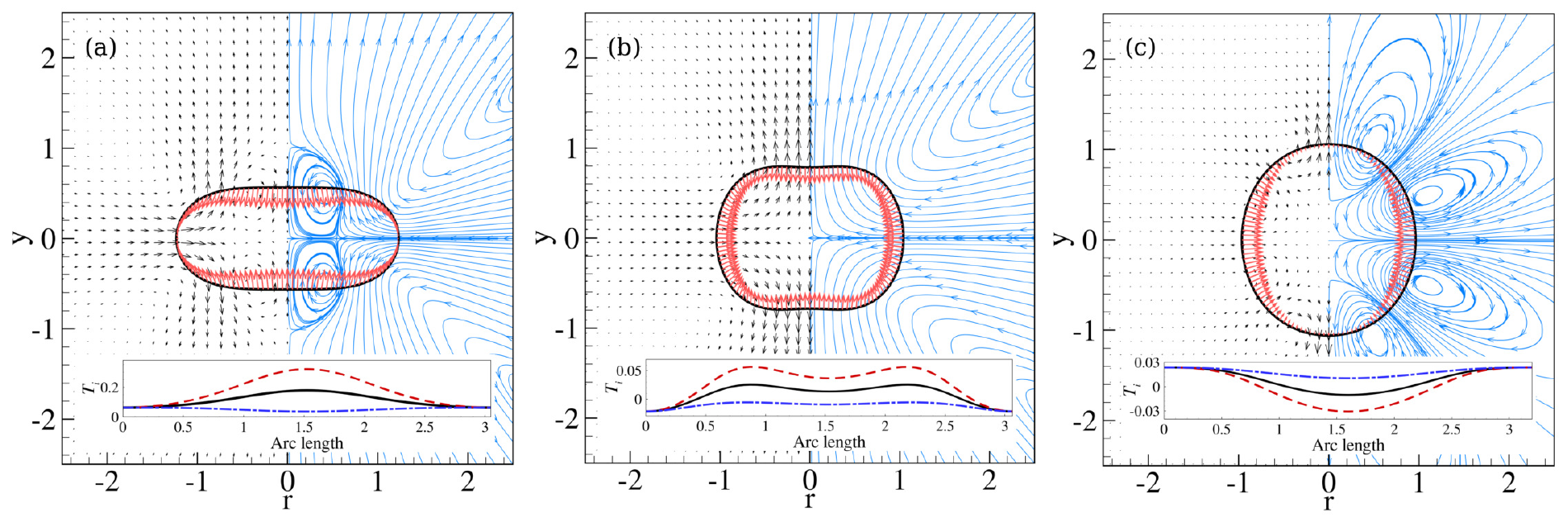}
  \caption{Electric stress (shown by arrows at the interface), streamlines (shown only at right half) and velocity profile (shown only at left half, magnitude represents the relative extent of flow) for the shapes shown in~\cref{fig:cp_shape}(a-c), respectively. The variation of meridional, $T_s$ (\textcolor{blue}{$\pmb{-\cdot-}$}), azimuthal, $T_\phi$ (\textcolor{red}{$\pmb{--}$}) and mean, $T_m$ (\textcolor{black}{$\pmb{\mi}$}) elastic tensions as a function of arc length are shown in insets. }
  \label{fig:cp}
 \end{figure*}

Figures~\ref{fig:cp_shape}a-d show the shapes observed in cylindrical-prolate (C-P) breathing mode in a single time-period. The shapes  shown correspond  to the measured deformation at the marker points on the curve in fig.~\ref{fig:cp_shape}e and the corresponding instantaneous electric fields are shown by marker points on the curve in~\cref{fig:cp_shape}f. Figures~\ref{fig:cp}a-c show the hydrodynamic flow, variation of electric and elastic stresses at the interface, for the shapes in figs.~\ref{fig:cp_shape}a-c, respectively. The shape in~\cref{fig:cp_shape}a is oblate since the outer medium is more conductive and  $t_{cap}^{-1}<\omega=0.02<t_{MW}^{-1}$. A tensile azimuthal elastic tension at the equator (characteristic of an oblate shape) develops~(inset of~\cref{fig:cp}a). The electric stresses are compressive at the poles~(\cref{fig:cp}a) assisting the formation of oblate shapes. The flow from the equator to the poles~(in~\cref{fig:cp}a) shows a tendency of the capsule to return to a relaxed shape~(\cref{fig:cp_shape}b). The high electric stress at the poles due to a highly nonlinear shape along with the compressive electric stress at the equator~(\cref{fig:cp}b) assist the formation of cylindrical shape~(\cref{fig:cp_shape}b) while admitting an almost spherical shape (not shown here) at  an intermediate time. The cylindrical shape is thus a manifestation of the high capillary number and corresponding interplay of high Maxwell stress and nonlinearity in shape (see figs.~\ref{fig:cp} and figs.~\ref{fig:strsqp0p1}). Towards the end of the cycle, the capsule assumes a prolate shape~(\cref{fig:cp_shape}c) which is due to the developed Maxwell stress of a completely charged capacitor, the compressive normal electric stress with a maximum at the equator~(\cref{fig:cp}c). The elastic tensions are accordingly generated, with the azimuthal tension becoming negative at the equator~(inset of~\cref{fig:cp}c). A tendency to attain oblate shape is then seen~(\cref{fig:cp_shape}d) as the field reverses and the membrane discharges to attain an electrostatic solution commensurate with the applied field.  

\begin{figure*}
  \centering
  \includegraphics[width=0.9\textwidth]{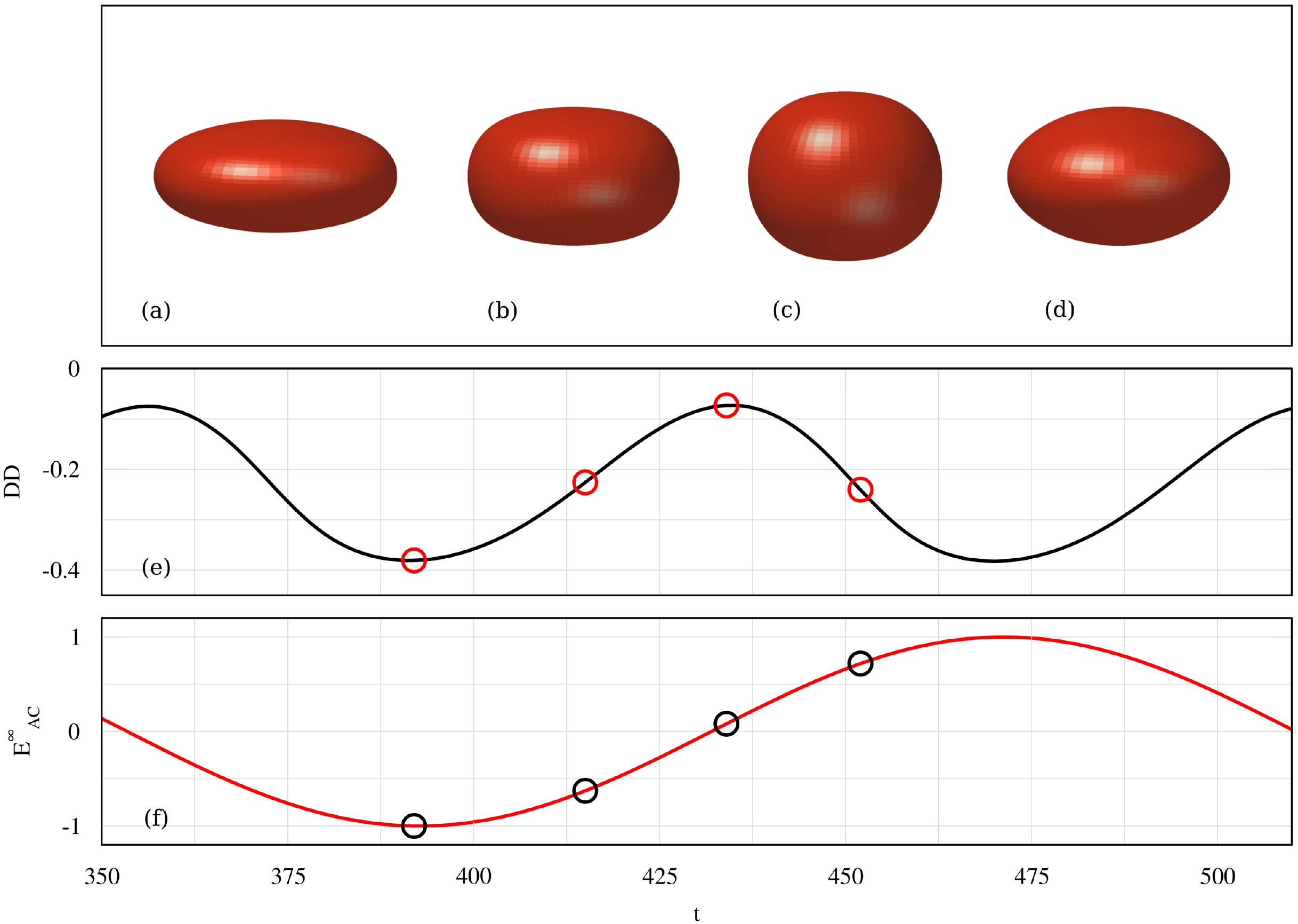}
  \caption{Cylinder-oblate (C-O) breathing of a capsule (a-d) at $Ca=0.4$ for $\omega=0.04$ and $\sigma_r=0.3$ considering $\epsilon_r=1$, $G_m=0$ and $C_m=50$. Shapes are corresponding to the marker points on the curves in (e) representing degree of deformation and in (f) representing applied electric field as the functions of time.}
  \label{fig:co_shape}
\end{figure*}

\begin{figure*}
  \centering
  \includegraphics[width=0.9\textwidth]{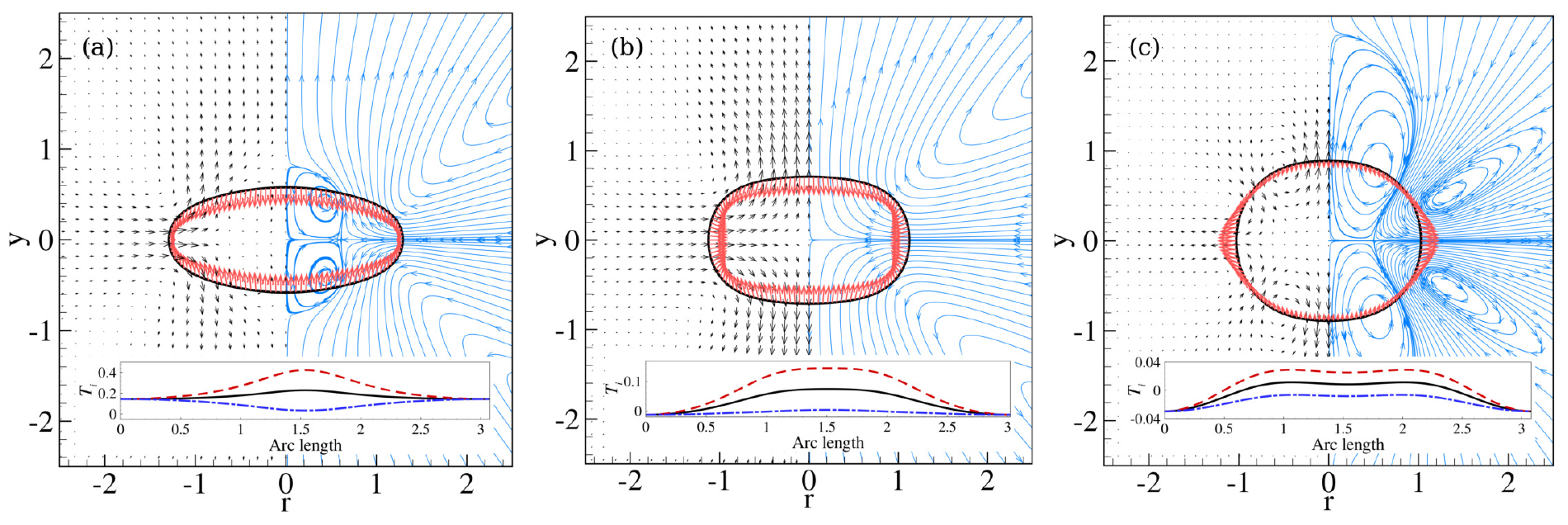}
  \caption{Electric stress (shown by arrows at the interface), streamlines (shown only at right half) and velocity profile (shown only at left half, magnitude represents the relative extent of flow) for the shapes shown in~\cref{fig:co_shape}(a-c), respectively. The variation of meridional, $T_s$ (\textcolor{blue}{$\pmb{-\cdot-}$}), azimuthal, $T_\phi$ (\textcolor{red}{$\pmb{--}$}) and mean, $T_m$ (\textcolor{black}{$\pmb{\mi}$}) elastic tensions as a function of arc length are shown in insets. }
  \label{fig:co}
\end{figure*}
  \begin{figure*}
  \centering
  \includegraphics[width=0.9\textwidth]{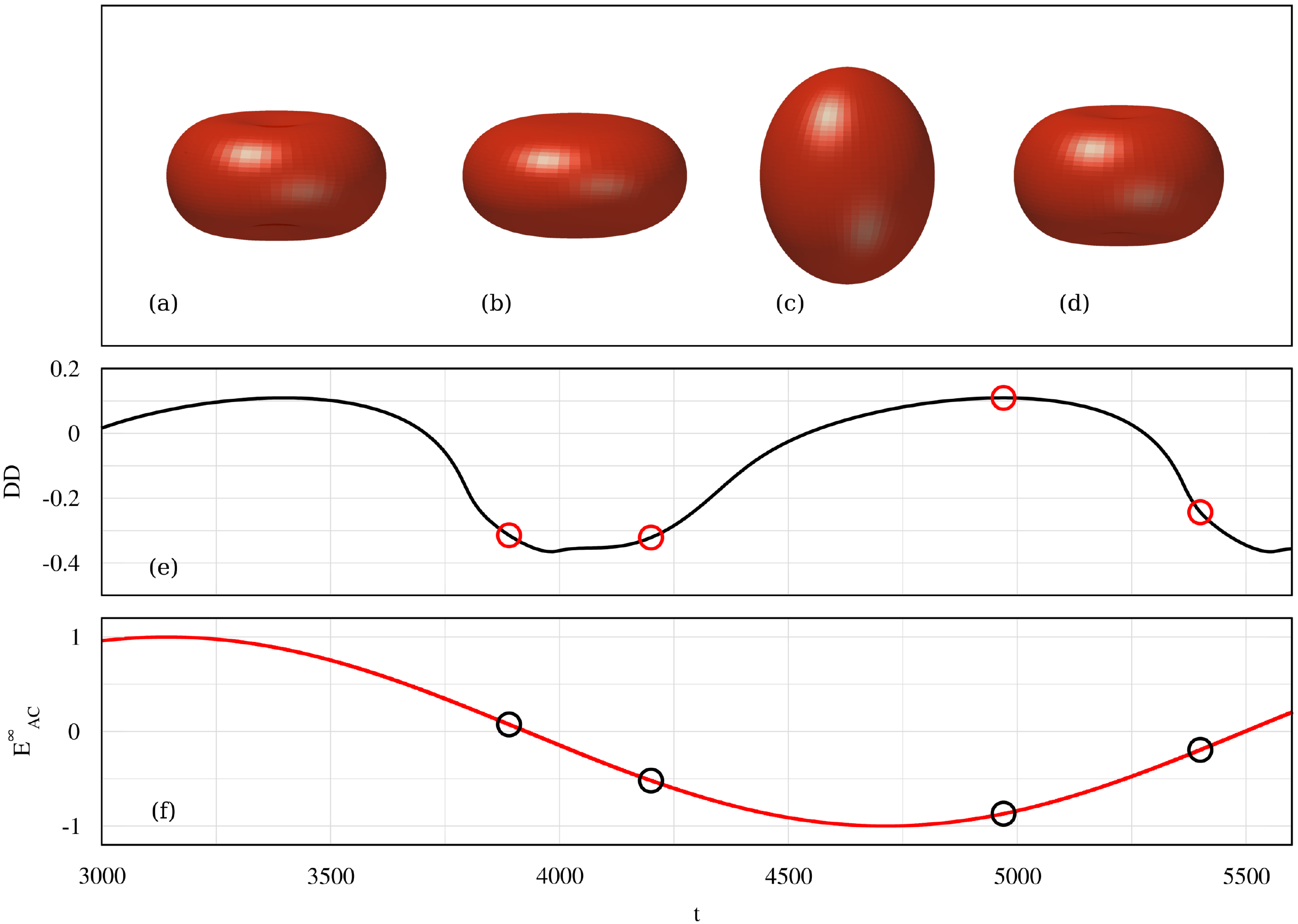}
  \caption{Biconcave-prolate (B-P) breathing of a capsule (a-d) at $Ca=0.4$ for $\omega=0.002$ and $\sigma_r=0.1$ considering $\epsilon_r=1$, $G_m=0$ and $C_m=50$. Shapes are corresponding to the marker points on the curves in (e) representing degree of deformation and in (f) representing applied electric field as the functions of time.}
  \label{fig:bp_shape}
\end{figure*}
    \begin{figure*}
  \centering
  \includegraphics[width=0.9\textwidth]{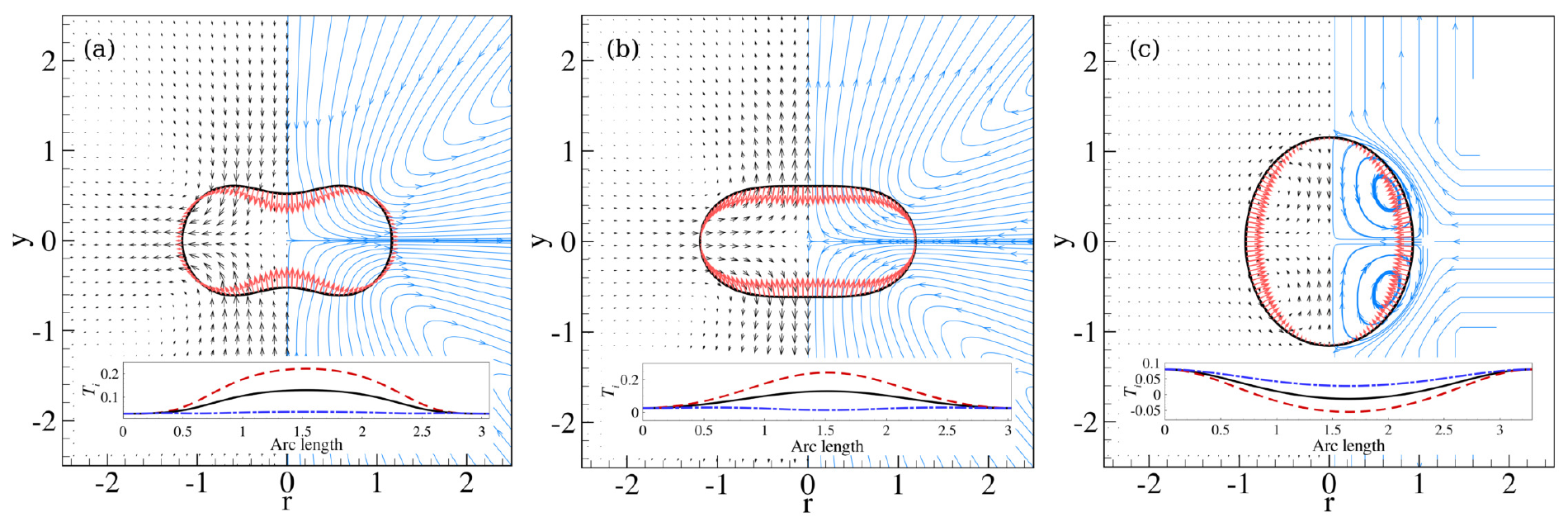}
  \caption{Electric stress (shown by arrows at the interface), streamlines (shown only at right half) and velocity profile (shown only at left half, magnitude represents the relative extent of flow) for the shapes shown in~\cref{fig:bp_shape}(a-c), respectively. The variation of meridional, $T_s$ (\textcolor{blue}{$\pmb{-\cdot-}$}), azimuthal, $T_\phi$ (\textcolor{red}{$\pmb{--}$}) and mean, $T_m$ (\textcolor{black}{$\pmb{\mi}$}) elastic tensions as a function of arc length are shown in insets. }
  \label{fig:bp}
\end{figure*}

\begin{figure}
\centering
  \includegraphics[width=0.45\textwidth, trim=0in 0in 0in 0in, clip]{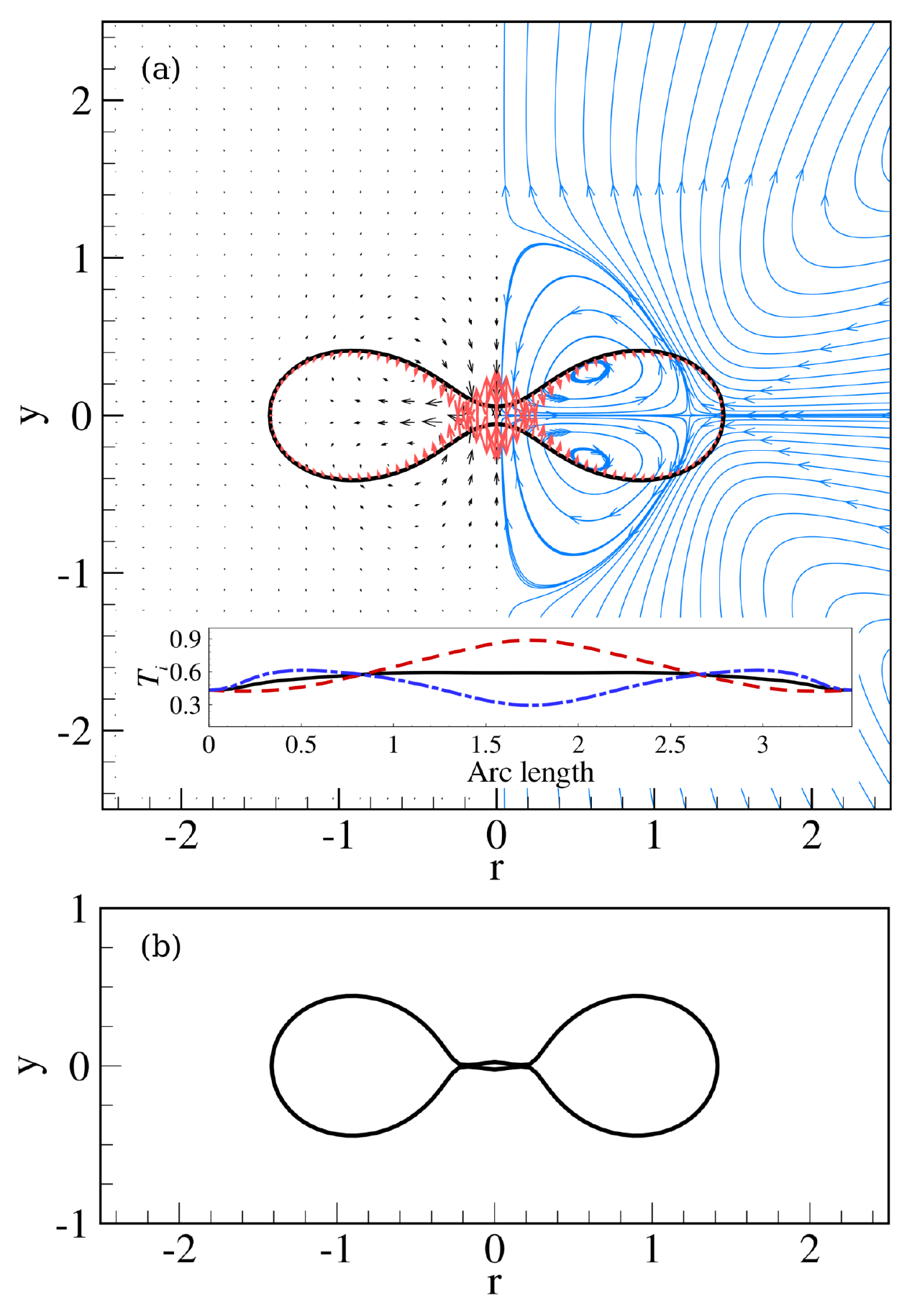}%
  \caption{Breakup of an elastic capsule at $Ca=0.4$ for $\omega=0.02$ and $\sigma_r=0.1$ considering $\epsilon_r=1$, $G_m=0$ and $C_m=50$. (a) Electric stress (shown by arrows at the interface), streamlines (shown only at right half) and velocity profile (shown only at left half, magnitude represents the relative extent of flow) for the shape at the onset of break up (coordinates are assigned for the shape of the capsule). The variation of meridional, $T_s$ (\textcolor{blue}{$\pmb{-\cdot-}$}), azimuthal, $T_\phi$ (\textcolor{red}{$\pmb{--}$}) and mean, $T_m$ (\textcolor{black}{$\pmb{\mi}$}) elastic tensions as a function of arc length are shown in insets. (b) Shape obtained from the numerical simulation just before the failure of the boundary integral code suggesting the breakup of the capsule through the merging of poles.}
  \label{fig:breakup}
\end{figure} 

For a slightly lower conductivity and a higher frequency (as compared to fig.~\ref{fig:cp_shape}), observed shapes during the deformation are shown in figs.~\ref{fig:co_shape}a-d. The corresponding  degree of deformation and the instantaneous applied electric fields are shown by the marker points on curves in~\cref{fig:co_shape}e and f. A capsule attains cylinder-oblate spheroid (C-O) breathing mode at $\sigma_r=0.3$ and $\omega=0.04$. Figures~\ref{fig:co}a-c show the state of electric stress, streamline, flow and elastic tensions (in insets) corresponding to figs.~\ref{fig:co_shape}a-c, respectively. In this case, on account of relatively higher frequency (higher than the inverse of the membrane charging time) and lower conductivity ratio (compared to the C-P breathing), the capsule never relaxes to a sphere or a prolate shape. Only oblate shapes are admitted with flow from the equator to poles~(figs.~\ref{fig:co}a and b) assisting the relaxation through the cylindrical shape~(\cref{fig:co_shape}b) to the oblate shape~(\cref{fig:co_shape}c). Pronounced tensile normal electric stresses are seen at the equator~(\cref{fig:co}c) (a mechanism similar to that discussed in the low capillary limit, see figs.~\ref{fig:strsqo0p1}) that are  responsible for the oblate shapes resulting in cylinder-oblate transition. 

At much lower frequency though, a biconcave-prolate (B-P) breathing mode is observed~(figs.~\ref{fig:bp_shape}a-d). 
The biconcave shape~(\cref{fig:bp_shape}a) is the result of increased compressive Maxwell stress at the poles~(\cref{fig:bp}a), owing to higher conductivity contrast. Since the frequency is low, a relaxation to prolate shape is seen at the end of the cycle ~(\cref{fig:bp_shape}c) via an intermediate oblate shape~(\cref{fig:bp_shape}b). This B-P breathing of capsule deformation is not observed at low capillary number as the electric stresses at the poles are insufficient (see figs.~\ref{fig:ca0p1bic}). 

\begin{figure}
\centering
  \includegraphics[width=0.45\textwidth]{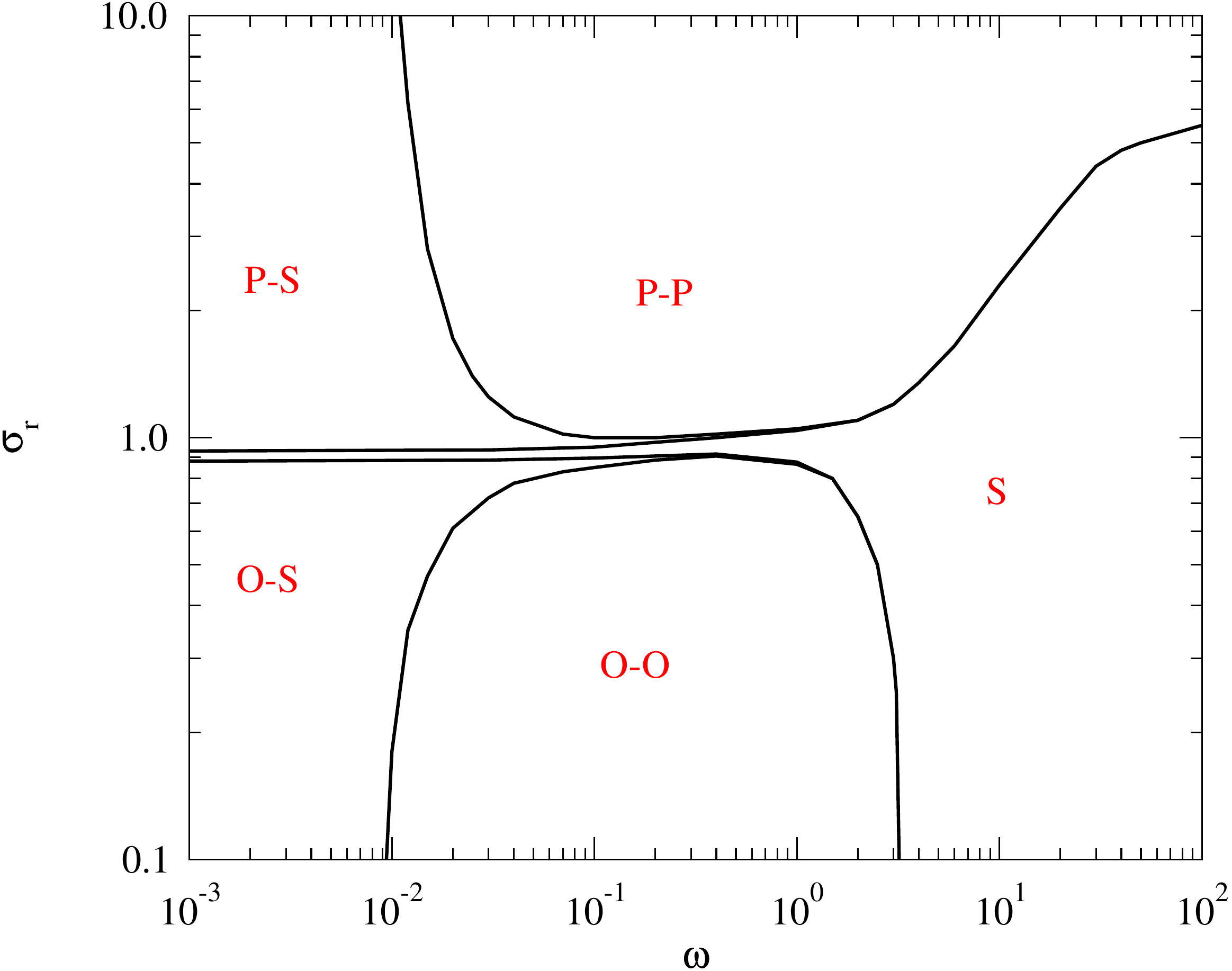}%
  \caption{Phase diagram of time-periodic breathing of capsule at $Ca=0.4$ considering $\epsilon_r=1$, $G_m=10$ and $C_m=50$ representing prolate-prolate breathing (P-P), prolate-sphere breathing (P-S), oblate-sphere breathing (O-S), oblate-oblate breathing (O-O) and undeformed sphere (S) zones of deformation.}
  \label{fig:gm10pda}
\end{figure} 

In this range of frequency ($5\times10^{-3}\lesssim\omega\lesssim2\times 10^{-2}$) and at very low conductivity ratios, on account of greater charging of the membrane (the capsule acts more like a leaky dielectric drop), a highly compressive normal Maxwell stresses develop at the poles~(e.g.,~\cref{fig:breakup}a).  When the elastic tractions can not balance the developed electric stress at the pole, a capsule breaks up through the merging of poles. This breakup zone is marked with X in the phase diagram~(\cref{fig:phase0p4}). \Cref{fig:breakup}a confirms that within the valid range of numerical calculation, the developed elastic tensions at the pole just before breakup are very high~(inset of~\cref{fig:breakup}a). It should be noted that when the collocation points at poles come very close (closer than the element size used for the computation), the numerical computation is no more valid. Therefore, the shape observed in~\cref{fig:breakup}b is actually a numerical artifact.\\ 

%

The modes of deformation of an elastic capsule with conducting membrane $(\hat{G}_m=10)$ in AC electric field for $Ca=0.4$ is represented in~\cref{fig:gm10pda}, keeping other parameters unchanged, i.e $\epsilon_r=1$ and $\hat{C}_m=50$. In this case, a capsule does not exhibit any complex breathing mode in an AC field for the range of frequency $10^{-3}\leq\omega\leq10^{2}$ and conductivity ratio $0.1\leq\sigma_r\leq10$. The P-P, P-S, O-S and O-O breathing modes are observed at low frequencies, but P-O breathing is not observed for this type of capsule. The high conductivity of the membrane in this case results in a capsule behaving like a leaky dielectric drop in a leaky dielectric medium. Thus, cylindrical or biconcave shapes are not observed since the stress remains highly compressive only at the poles, lacking any contribution from the equator. 

 \section{Conclusions} 
A phase diagram is constructed using boundary integral method and analytical theory at a low capillary number. It is found that the time-averaged deformation depends upon the frequency of the applied electric field and its value relative to the $t_{MW}^{-1}$ and $t_{cap}^{-1}$. The time-periodic deformation shows breathing modes characterized as P-P, P-O, and O-O at a low capillary number. A high capillary number causes further bifurcation to highly nonlinear C-P, C-O, and B-P modes of deformation, attributed to strong coupling between charge density and electrostatics. \\

The proposed experiments to observe the theoretical results presented in this work could be on a thin Ovalbumin or human serum albumin (HSA) capsules of radius $a\simeq 500\ \mu m$ with elasticity $E_s\simeq 0.1\ N/m$~\cite{chu11}, in an external fluid medium  with conductivity $\sigma_e\simeq 0.1\ mS/m$. In these experimental conditions, the typical field strength $E_0<5\ kV/cm$ and the frequencies in the range of $[100\ Hz-10\ MHz]$ should show the phenomenon described in this work.  

This work is the simplest system that can demonstrate a strong shape-electrostatic-elastic-hydrodynamic coupling, which is important in electrohydrodynamics of soft matter. Findings of this work can form the basis for understanding large electro-elasto-hydrodynamics of more complicated soft matter systems, such as RBC, which will be the project for future communication.\\

\section*{Acknowledgment}
Authors would like to thank the Department of Science and Technology (DST), Govt. of India, for financial support.

\appendix
\section{Analytical theory for electrohydrodynamic deformation of a capsule in the limit of small deformation}\label{sec:antheory}
To seek an analytical solution in the small deformation limit, a spherical coordinate system is considered $(r,\theta,\phi)$, where $r$ is the radial measurement from the center, $\theta$ is the measurement of meridional angle and $\phi$ is the measurement of azimuthal angle. In the axisymmetric analysis, assuming $\theta=0$ as the axis of symmetry, variations of quantities are independent of azimuthal angle. 
 
The nondimensional potential due to the applied electric field in spherical coordinate system is expressed by
\begin{equation}
 \phi^\infty=-\cos(\omega t)r\cos\theta
\end{equation}
which can be written as
\begin{equation}
 \bar{\phi}^\infty=-\frac{1}{2}\left(e^{j\omega t}+e^{-j\omega t}\right)r\cos\theta.
\end{equation}
Henceforth, a variable $\bar{x}$ is represented as a function of $x$ and its conjugate $x^*$ as
\begin{equation}
 \bar{x}=\frac{1}{2}(xe^{j\omega t}+xe^{-j\omega t}).
\end{equation}
Potential for the interior and exterior of the capsule satisfy the Laplace's equation,
\begin{equation}
 \frac{1}{r^2}\frac{r^2\frac{\partial \phi_{i,e}}{\partial r}}{\partial r}+\frac{1}{r^2\sin\theta}\frac{\partial\left(\sin\theta\frac{\partial\phi_{i,e}}{\partial\theta}\right)}{\partial \theta}=0
\end{equation}
where
\begin{align}\label{eq:phiie}
  \bar{\phi}_{i,e}=\frac{1}{2}\left[\phi_{i,e}e^{j\omega t}+\phi_{i,e}^*e^{-j\omega t}\right]
\end{align}
Internal and external potentials are expressed by spherical harmonics,
\begin{align}
 \phi_e &= \phi^\infty+\frac{A_1}{r^2}P_1(\cos\theta)\label{eq:phie}\\
 \phi_i &= A_2rP_1(\cos\theta)\label{eq:phii},
\end{align}
where, $A_1$ and $A_2$ are arbitrary constants, and $P_1(\cos\theta)$ is the first order Legendre polynomial of $\cos\theta$. 
Transmembrane potential is defined as 
\begin{equation}\label{eq:tmpot}
 \phi_m=\phi_{amp}\cos(\theta)=\phi_i-\phi_e
\end{equation}
Current continuity in the normal direction is given by
\begin{equation}\label{eq:current}
 \sigma_r E_{n,i}+\epsilon_r\frac{dE_{n,i}}{dt}=E_{n,e}+\frac{dE_{n,e}}{dt}=\hat C_m \frac{d \phi_m}{dt}+ \hat G_m \phi_m
\end{equation}
Coefficients $A_1$, $A_2$ and the transmembrane potential, $\phi_m$, can be obtained from the solutions of~\cref{eq:phie} and~\cref{eq:phii} using the definition of transmembrane potential~(\cref{eq:tmpot}) and the current continuity~(\cref{eq:current}). The solution for the transmembrane potential of a spherical capsule with a nonconducting membrane is obtained as 
\begin{widetext}
\begin{equation}
 \phi_m = 3\cos\theta\frac{2(1+\omega^2)(\omega^2\epsilon_r^2+\sigma_r^2)+\omega^2[\epsilon_r\{2+\omega^2(2+\epsilon_r))+\sigma_r^2\}\hat{C}_m]\cos(\omega t)+\omega\{\sigma_r(2+\sigma_r)+\omega^2(\epsilon_r^2+2\sigma_r)\}\hat{C}_m\sin(\omega t)}{
 4\sigma_r^2+\omega^2\{4(\epsilon_r^2+\sigma_r^2)+4(2\epsilon_r+\sigma_r^2)\hat{C}_m+(2+\sigma_r)^2\hat{C}_m^2\}+\omega^4\{2\hat{C}_m+\epsilon_r(2+\hat{C}_m)\}^2},\label{eq:phimexp}
\end{equation}
\end{widetext}
with a phase lag 
\begin{equation}
 \alpha=-\arctan\left[\frac{\omega(\sigma_r(2+\sigma_r)+\omega^2(\epsilon_r^2+2\sigma_r))\hat{C}_m}{2(1+\omega^2)(\omega^2\epsilon_r^2+\sigma_r^2)+\omega^2[\epsilon_r\{2+\omega^2(2+\epsilon_r)\}+\sigma_r^2]\hat{C}_m}\right].
\end{equation}
Electric fields in $r$ and $\theta$ directions at the interface are determined as
\begin{subequations}
\label{eq:eret}
\begin{eqnarray}
 E_{r,i}     &=&-\frac{\partial \phi_i}{\partial r}\\
 E_{\theta,i}&=&-\frac{1}{r}\frac{\partial \phi_i}{\partial \theta}\\
 E_{r,e}     &=&-\frac{\partial \phi_e}{\partial r}\\
 E_{\theta,e}&=&-\frac{1}{r}\frac{\partial \phi_E}{\partial \theta}.
 \end{eqnarray}
 \end{subequations}

The Maxwell electric stress tensor is given by $\tilde{\bf T}^E=\epsilon(\tilde{\bf E}\tilde{\bf E}-\frac{1}{2}\tilde{E}^2{\bf I})$, where $\tilde{E}^2$ is the inner product of the field, and ${\bf I}$ is the identity tensor. The normal and tangential components of the electric traction can be obtained by
\begin{subequations}
\begin{eqnarray}
 \tilde{f}_r &=&{\bf n}\cdot \tilde{\bf T}^E\cdot {\bf n}=\frac{1}{2}\epsilon\epsilon_0(\tilde{E}_r^2-\tilde{E}_\theta^2)\\
 \tilde{f}_\theta &=&{\bf t}\cdot \tilde{\bf T}^E\cdot {\bf n}=\epsilon\epsilon_0 \tilde{E}_r\tilde{E}_\theta
\end{eqnarray}
\end{subequations}
such that
$\tilde{\bf f}^E=\tilde{f}_r {\bf e}_r+\tilde{f}_\theta {\bf e}_\theta$ is the total electric traction acting at the interface. For an undeformed sphere, normal vector ${\bf n}={\bf e}_r$  and the tangent vector ${\bf }t={\bf e}_\theta$.
Using eqs.~\ref{eq:eret} with the~\cref{eq:phiie}, dimensionless tractions can be expressed as
\begin{widetext}
\begin{subequations}
\begin{eqnarray}
   \bar{f_r} &=& \frac{1}{8}\left[2E_{re}E^*_{re}-2E_{\theta e}E^*_{\theta e}+(E_{re}^2-E_{\theta e}^2)e^{2j\omega t}+(E^*_{re}-E^*_{\theta e})e^{-2j\omega t}\right]\nonumber\\
 &&-\frac{\epsilon_r}{8}\left[2E_{ri}E^*_{ri}-2E_{\theta i}E^*_{\theta i}+(E_{ri}^2-E_{\theta i}^2)e^{2j\omega t}+(E^*_{ri}-E^*_{\theta i})e^{-2j\omega t}\right]\\
 \bar{f_\theta} &=& \frac{1}{4}\left[(E_{re}E^*_{\theta e}+E^*_{r e}E_{\theta e})+(E_{re}E_{\theta e}e^{2j\omega t}+E^*_{re}E^*_{\theta e}e^{-2j\omega t})\right]\nonumber\\
 &&-\frac{\epsilon_r}{4}\left[(E_{ri}E^*_{\theta i}+E^*_{r i}E_{\theta i})+(E_{ri}E_{\theta i}e^{2j\omega t}+E^*_{ri}E^*_{\theta i}e^{-2j\omega t})\right].
\end{eqnarray}
\end{subequations}
\end{widetext}
Tractions can also be expressed as
\begin{subequations}
\begin{eqnarray}
 f_r &=&f_{rs}+\frac{1}{2}\left[f_{rt}e^{2j\omega t}+f^*_{rt}e^{-2j\omega t}\right]\\
 f_\theta &=&f_{\theta s}+\frac{1}{2}\left[f_{\theta t}e^{2j\omega t}+f^*_{\theta t}e^{-2j\omega t}\right],
\end{eqnarray}
\end{subequations}
where subscripts $s$ and $t$ represent time-averaged and time-periodic parts. Time-averaged electric tractions are obtained as
\begin{widetext}
\begin{subequations}
\begin{eqnarray}
 f_{rs} &=&\frac{1}{4}[E_{re}E^*_{re}-E_{\theta e}E^*_{\theta e}]-\frac{\epsilon_r}{4}[E_{ri}E^*_{ri}-E_{\theta i}E^*_{\theta i}],\\
 f_{\theta s} &=&\frac{1}{4}[E_{re}E^*_{\theta e}+E^*_{re}E_{\theta e}]-\frac{\epsilon_r}{4}[E_{ri}E^*_{\theta i}+E^*_{ri}E_{\theta i}]
\end{eqnarray}
\end{subequations}
\end{widetext}
and time-periodic electric tractions are 
\begin{subequations}
\begin{eqnarray}
f_{rt}&=&\frac{1}{4}\left[(E_{re}^2-E_{\theta e}^2)-\epsilon_r(E_{ri}^2-E_{\theta i}^2)\right],\\
f_{\theta t}&=&\frac{1}{2}\left[E_{re}E_{\theta e}-\epsilon_rE_{ri}E_{\theta i}\right],
\end{eqnarray}
\end{subequations}
where $f_{rs}$, $f_{\theta s}$ are real and $f_{rt}$, $f_{\theta t}$ are complex quantities. 

Calculation of elastic traction, hydrodynamics formulation, and solution procedure are followed as reported in our earlier article \cite{rt16}. In general, the expressions for the degree of deformation as a function of frequency and capillary number are very complex. For a nonconducting $(G_m=0)$ membrane, the time-averaged degree of deformation is obtained as
\begin{widetext}
\begin{equation}
 DD_s=\frac{9}{32}Ca\frac{\Big(5(\epsilon_r^2+\sigma_r^2)+2\hat{C}_m(5\epsilon_r+3\sigma_r^2)+\hat{C}_m^2\{5-16\epsilon_r+\sigma_r(6+5\sigma_r)\}\Big)\omega^2}
  {4\sigma_r^2+(4(\hat{C}_m+\epsilon_r)^2+4\hat{C}_m^2\sigma_r+(2+\hat{C}_m)^2\sigma_r^2)\omega^2+(2\epsilon_r+\hat{C}_m(2+\epsilon_r))^2\omega^4}\label{eq:ddf},
\end{equation}
\end{widetext}
and the time-periodic degree of deformation for a special case considering $G_m=0$, $C_m=50$, $\epsilon_r=1$ is obtained as
\begin{widetext}
\begin{eqnarray}
 DD_t&=&\frac{45Ca}{32}\Bigg[\sigma_r^2+6\sigma_r\Big(17+11\sigma_r\Big)j\omega
 +\Big\{5399-\sigma_r(3732+2869\sigma_r)\Big\}\omega^2-
 2\Big\{\sigma_r(12319+5152\sigma_r)-16167\Big\}j\omega^3\nonumber\\
 &&+\Big(36008\sigma_r-33731\Big)\omega^4+704j\omega^5\Bigg]\Bigg/
 \Bigg[\Big\{35\omega(10\omega-3j)-4\Big\}
 \Big\{(51+76j\omega)\omega+\sigma_r(26\omega-j)\Big\}^2\Bigg].
\end{eqnarray}
\end{widetext}
The deformation of an elastic capsule, in the limit of small deformation, can be obtained from
\begin{equation}
 DD=DD_s+\left[\frac{DD_te^{2j\omega t}+DD_t^*e^{-2j\omega t}}{2}\right],
\end{equation}
where $DD_t^*$ is the complex conjugate of $DD_t$.

\bibliography{rsc} 
\bibliographystyle{plainnat} 
\pagebreak
\section{Supplementary information}
\begin{figure}
\centering
\includegraphics[width=0.45\textwidth]{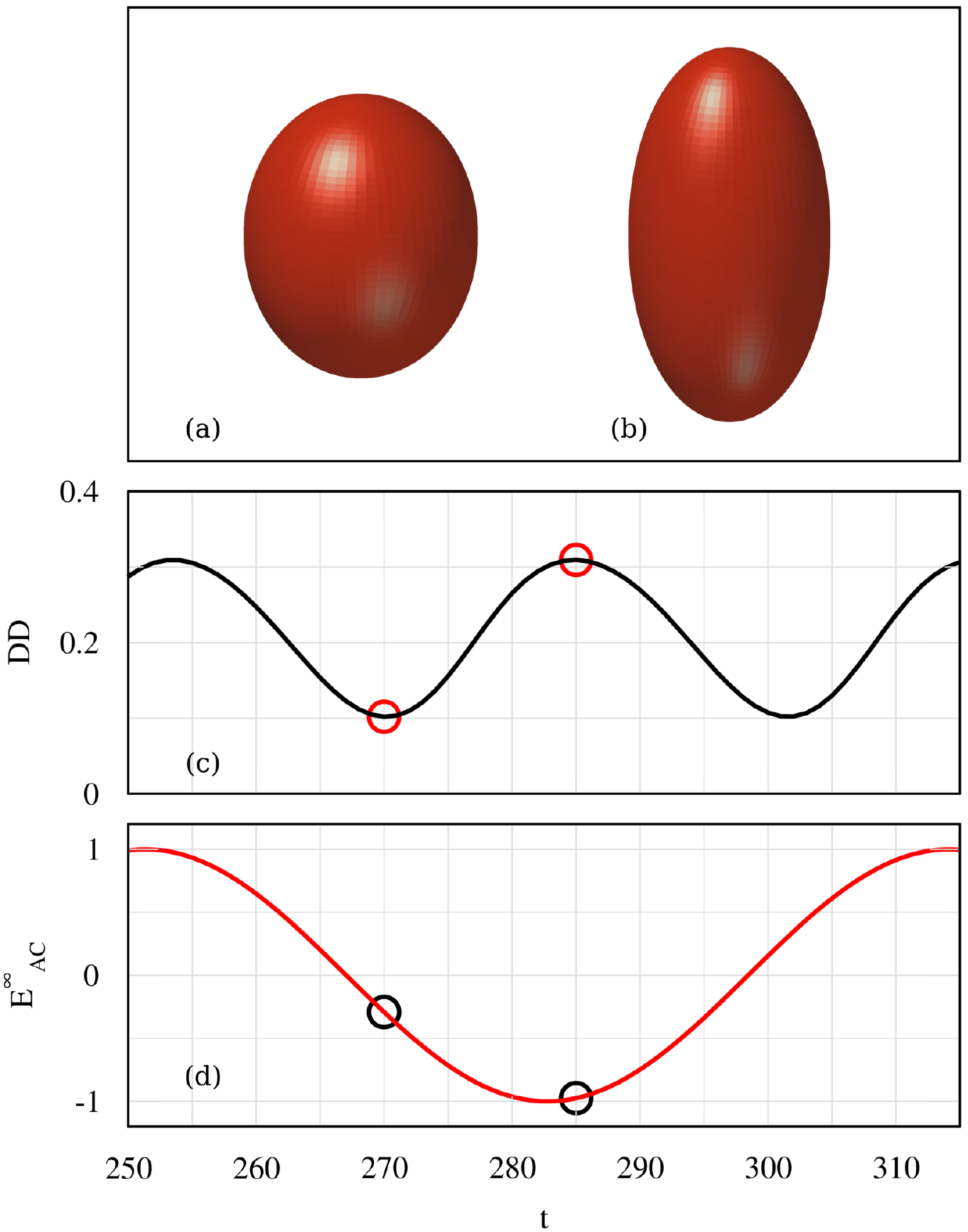}%
\caption{P-P breathing of a capsule at $Ca=0.4$ for $\omega=0.1$ and $\sigma_r=5$, considering $\epsilon_r=1$, $G_m=0$ and $C_m=50$. Shapes are corresponding to the marker points on the curves in (c) representing the degree of deformation and in (d) representing applied electric field as the functions of time.}
  \label{fig:pp}
\end{figure}

 The prolate-prolate breathing (P-P) of an elastic capsule at $\omega=0.1$ and $\sigma_r=5$, is shown in figs.~\ref{fig:pp}a and b. The degrees of deformation of the shapes shown in figs.~\ref{fig:pp}a and b are shown with the marker points on the curve in~\cref{fig:pp}c and the corresponding instantaneous applied electric field is shown in~\cref{fig:pp}d. This example of P-P breathing is corresponding to the zone presented in the high capillary number phase diagram in~\cref{fig:phase0p4}.

\begin{figure}
\centering
  \includegraphics[width=0.45\textwidth]{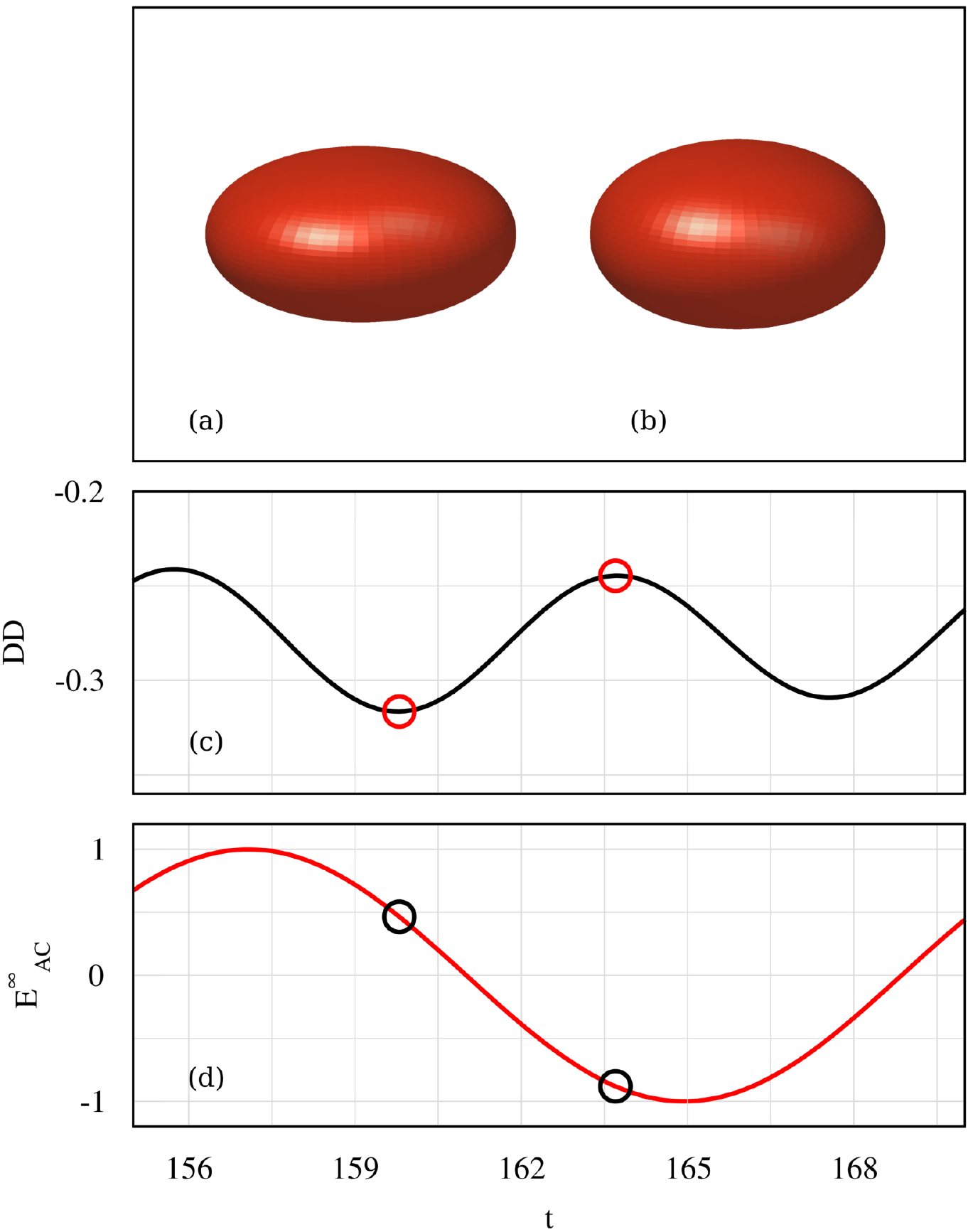}
\caption{O-O breathing of a capsule at $Ca=0.4$ for $\omega=0.4$ and $\sigma_r=0.1$, considering $\epsilon_r=1$, $G_m=0$ and $C_m=50$. Shapes are corresponding to the marker points on the curves in (c) representing the degree of deformation and in (d) representing applied electric field as the functions of time.}
  \label{fig:oo}
\end{figure}

Similar to the case of P-P breathing, a capsule with the $\sigma_r$ in the O-O zone, experience oblate-oblate breathing in AC electric field. As an example, the shapes at the minima and maxima of the electric stress in the O-O breathing at $Ca=0.4$ for $\omega=0.4$ and $\sigma_r=0.1$ are shown in figs.~\ref{fig:oo}a and b. The degree of deformation of these shapes are shown by the marker points on the deformation curve~(\cref{fig:oo}b) and the corresponding instantaneous electric fields are shown by the marker points on the curve in~\cref{fig:oo}c. 
  
\begin{figure}
\centering
  \includegraphics[width=0.45\textwidth]{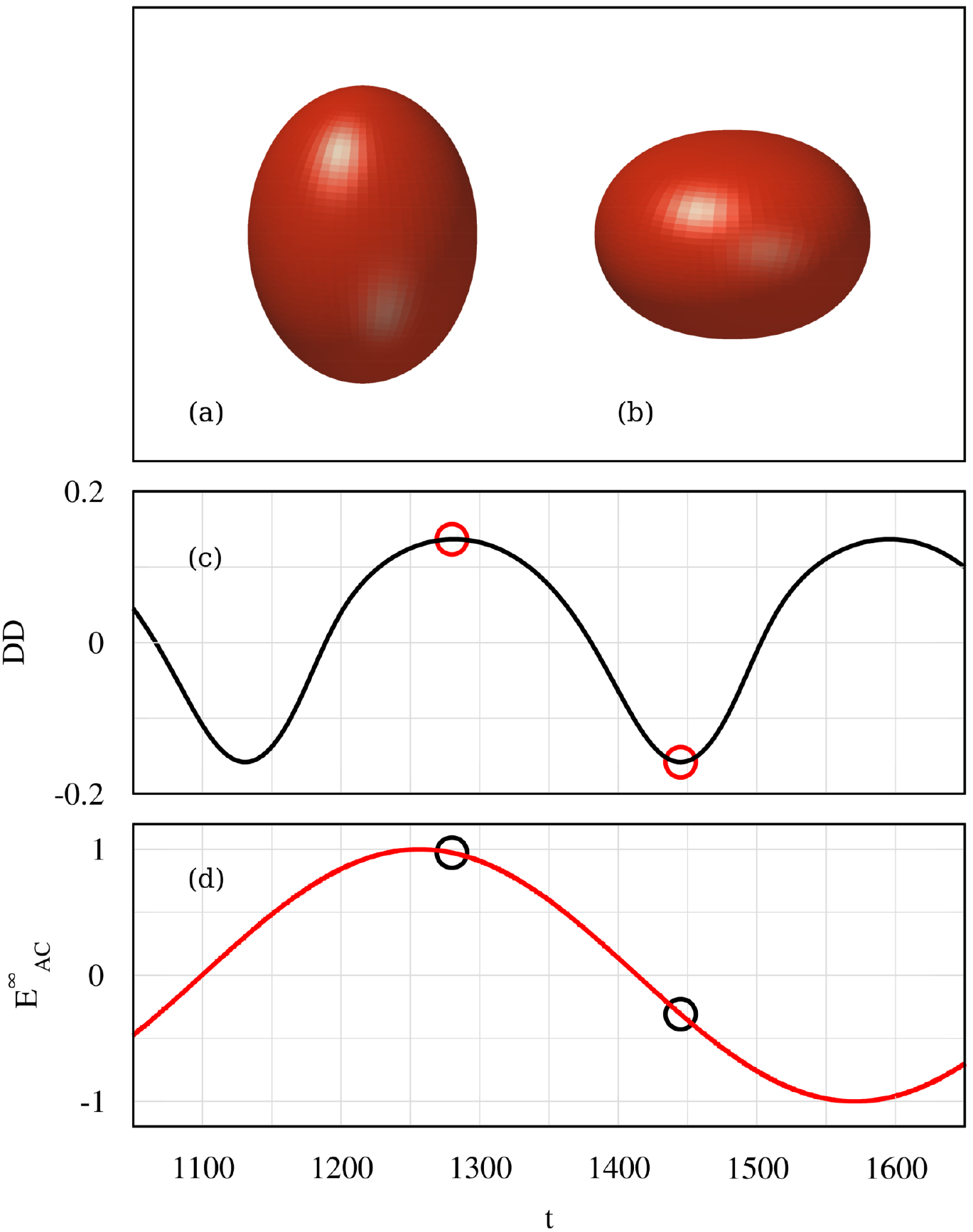}%
\caption{P-O breathing of a capsule at $Ca=0.4$ for $\omega=0.01$ and $\sigma_r=0.8$, considering $\epsilon_r=1$, $G_m=0$ and $C_m=50$. Shapes are corresponding to the marker points on the curves in (c) representing the degree of deformation and in (d) representing applied electric field as the functions of time.}
  \label{fig:po}
\end{figure}

At very low frequency and conductivity ratio of O-O zone, (for an example, at $Ca=0.4$ for $\omega=0.01$ and $0.8$) the extreme shapes in P-O breathing are shown in figs.~\ref{fig:po}a and b which are corresponding to the deformations, shown by the marker point on the curve in~\cref{fig:po}c and the corresponding instantaneous electric fields are shown by marker points on the curve in~(\cref{fig:po})d.


\begin{figure*}
 \begin{center}
  \includegraphics[width=1\textwidth, trim=0.0in 0.0in 0.0in 0.0in, clip]{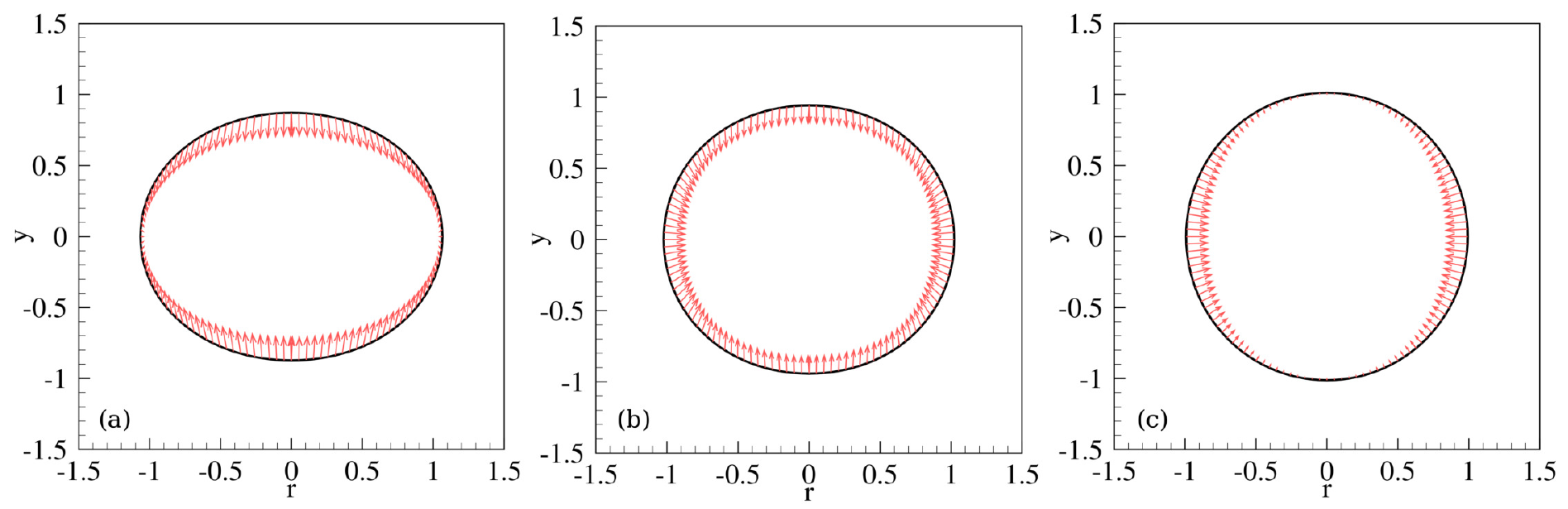}
\caption{In the limit of small capillary number ($Ca=0.1$) for $\omega=0.02$ and $\sigma_r=0.4$ (corresponding to the C-P breathing at high capillary number) considering $\epsilon_r=1$, $G_m=0$ and $C_m=50$, the electric stresses (arrows at the interface) are shown for the shapes observed during O-S breathing.}
\label{fig:strsqp0p1}
\end{center}
\end{figure*}

\begin{figure*}
 \begin{center}
  \includegraphics[width=1\textwidth, trim=0.0in 0.0in 0.0in 0.0in, clip]{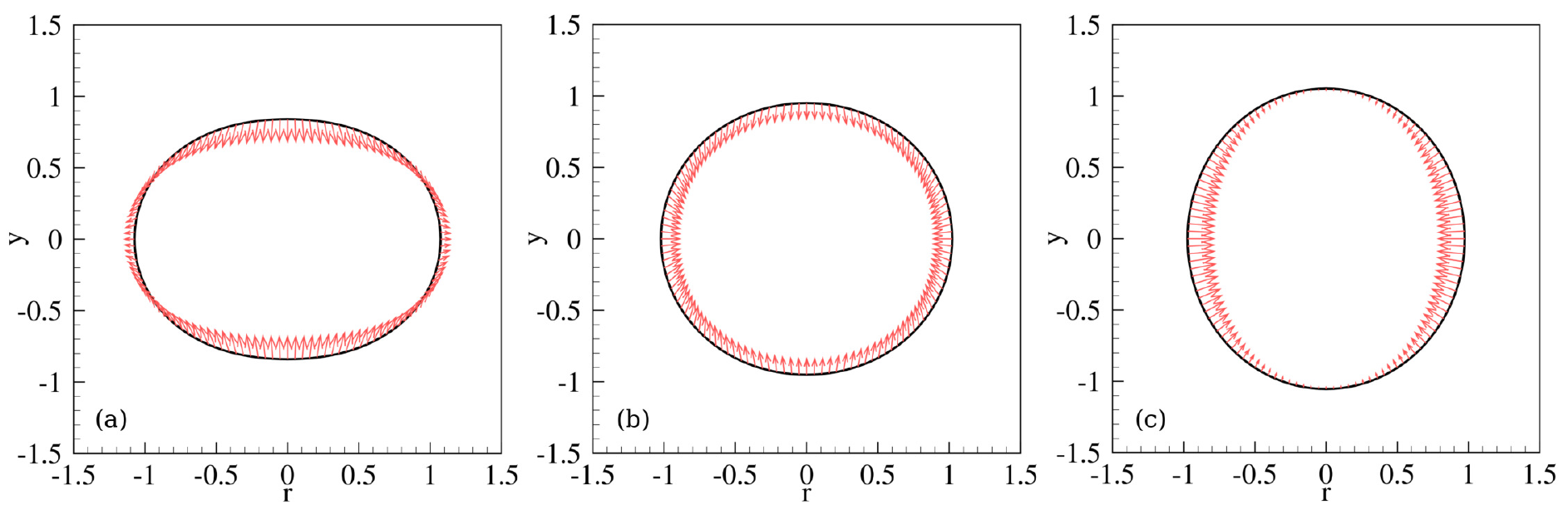}
\caption{In the limit of small capillary number ($Ca=0.1$) for $\omega=0.002$ and $\sigma_r=0.1$ (corresponding to the B-P breathing at high capillary number) considering $\epsilon_r=1$, $G_m=0$ and $C_m=50$, the electric stresses (arrows at the interface) are shown for the shapes observed during P-O breathing. }
\label{fig:ca0p1bic}
\end{center}
\end{figure*}

 At a small capillary number ($Ca=0.1$) the distribution of electric stress at the interface for $\omega=0.02$ and $\sigma_r=0.4$ is shown in~\cref{fig:strsqp0p1} undergoing O-S breathing and the same is shown for $\omega=0.002$ and $\sigma_r=0.1$ in~\cref{fig:ca0p1bic} representing P-O breathing of a capsule.  \Cref{fig:strsqp0p1,fig:ca0p1bic} are shown at the combinations of $\omega$ and $\sigma_r$ which are corresponding to the C-P and B-P breathing modes, respectively, at a high capillary number.
 
\end{document}